\documentclass[twocolumn,showpacs,preprintnumbers,amsmath,amssymb,nofootinbib]
{revtex4-1}

\usepackage{graphicx}
\usepackage{dcolumn}
\usepackage{bm}
\usepackage{verbatim} 
\usepackage{epsfig}
\usepackage{amsmath}
\usepackage{amsthm}
\usepackage{subfigure}

\begin{document}

\title{
Octupole Focusing Relativistic Self-Magnetometer Electric 
Storage Ring ``Bottle''}
\author{
Richard Talman \\
Laboratory of Elementary-Particle Physics, Ithaca, NY, USA \\
Cornell University \\
}
\author{
John Talman \\
UAL Consultants, Ithaca, NY, USA
}

\date{\today}

\begin{abstract}
A method proposed for measuring the electric dipole moment (EDM)
of a charged fundamental particle such as the proton, is to measure
the spin precession caused by a radial electric bend field $E_r$, 
acting on the EDMs of frozen spin polarized protons circulating in an
all-electric storage ring. The dominant systematic error limiting
such a measurement comes from spurious spin precession caused by 
unintentional and unknown average radial magnetic field $B_r$
acting on the (vastly larger) magnetic dipole moments (MDM) of 
the protons. Along with taking extreme magnetic shielding measures, 
the best protection against this systematic error is to use the storage 
ring itself, as a ``self-magnetometer''; the exact magnetic field
average $\langle B_r\rangle$ that produces systematic EDM error, is
nulled to exquisite precision by orbit position control.

The self-magnetometry sensitivity depends inversely on the 
restoring force with which the storage ring opposes the magnetic field
$\langle B_r\rangle$. By using octupole rather than quadrupole
focusing (which the name ``bottle'', copied from low energy
physics, is intended to convey) the restoring force can be vanishingly 
small for small amplitude vertical betatron-like motion while, at the 
same time, being strong enough at large amplitudes to keep all particles 
captured. This greatly enhances the magnetometer sensitivity.

In a purely electric ring clockwise (CW) and counter-clockwise (CCW) 
orbits would be identical, irrespective of ring positioning and powering
errors. In the absence of magnetic fields this symmetry is 
guaranteed by time reversal invariance (T). However, any 
average radial magnetic field error $\langle\Delta B_r\rangle$ causes
a vertical orbit shift between CW and CCW beams. 

Self-magnetometry
measures this shift, enabling its cancellation. For the octupole-only
ring proposed here the accuracy of magnetic field control is
$\langle\Delta B_r\rangle\approx \pm 3\times10^{-16}\,$Tesla.
This is small enough to reduce the systematic error in the proton 
EDM measurement into a range where realistically small deviations 
from standard model predictions can be measured.

Though novel, the theoretical analysis given here for relativistic 
bottles, either magnetic or electric, is elementary, and their
behavior is predicted to be entirely satisfactory. For particles
other than p and e, combined magnetic plus electic rings are needed, but
the same sefl-magnetometry should be applicable.
\end{abstract}


\maketitle

\tableofcontents

\section{Introduction}
For a fundamental particle such as the proton to have a non-zero
electric dipole moment (EDM) would violate both time reversal (T) 
and parity (P) symmetries. Both of these symmetries are, in fact, 
violated in the standard model, but too weakly to account for the observed 
matter-antimatter imbalance in the universe. This motivates
measuring the proton EDM. This will require trapping an intense polarized 
proton beam in a ``frozen spin'' all-electric storage ring ``trap''. 
Even in an electric field the spin precession of a moving proton is 
dominated by the motional magnetic field in the proton's rest
frame, acting on the magnetic dipole moment (MDM). But, the EDM and MDM 
precessions can be distinguished, because their precession axes are orthogonal. 

An ideal configuration for precision measurement of elementary particle 
parameters, 
such as magnetic dipole moments, would maintain the particles in a wall-free 
configuration stabilized by the electrostatic interaction of the charges.
Earnshaw's theorem proves this to be impossible. Nevertheless, by introducing 
RF cavities, lasers, etc., not covered by the theorem, 
it has been possible to store 
low energy particles indefinitely in various ``bottles'' or ``traps''. 
Relativistic charged particles can even be stored, wall-free, 
in storage rings, which rely on 
quadrupoles, solenoids, field gradients, or pole-edge rotation, 
to provide the ``linear'' focusing traditionally thought to be necessary 
to keep the particles captured.

This paper proposes an electric ``bottle'' capable of storing an intense beam 
of relativistic protons (or electrons, or other charged particles). 
Except for RF cavity (required to provide longitudinal, bunched beam, 
stability) and occasional brief ``reflections'' from 
octupole fields, the particles survive indefinitely, most of the time in 
nearly uniform fields. This proposal has been motivated primarily by the 
requirements for measuring the proton electric dipole moment (EDM) or, 
more particularly, by the 
need to suppress the most important systematic error limiting that measurement. 
The reflection mechanism resembles the end reflections in (non-relativistic) 
``magnetic mirror machines'' developed independently by Post 
in the U.S.A. and Budker in the U.S.S.R. But, unlike those devices, 
small amplitude particles cannot escape the proposed trap.

Various designs have been proposed for measuring the electric dipole
moments of charged particles, especially the 
proton\cite{FreqDomainEDM}\cite{BNLproposal}
and electron, in all-electric, frozen spin, storage rings. 
For an all-electric ring any non-zero EDM will cause spin 
precession about an axis forbidden by P and T symmetry. 
The overwhelmingly most serious 
background will be due to the presence of radial magnetic field $\Delta B_r$.
Here the $\Delta$ connotes that the field would ideally be zero, and the $r$ 
is the radially outward coordinate. (Following standard accelerator terminology,
the $E_r$ field will, where appropriate, be expressed as $E_x$, 
where $x$ is the radial
component in a local, Frenet coordinate system.) The field $\Delta B_x$, acting 
on the particle magnetic moment, causes precession that exactly mimicks the 
precession caused by the primary bending electric field, $E_x$, acting on 
the (vastly weaker) EDM. Even with $\Delta B_x$ reduced to the extent possible,
it will be challenging to reduce the systematic error to a value small enough 
to provide a serious test of the standard model (which predicts a value 
$d_p$ less than $10^{-30}\,$e-cm, for the proton EDM).

Reducing $\Delta B_x$ is not the only motivation for all-electric bending.
Electric bending also permits reversing the beam 
direction of circulation without changing 
any ring parameters. With vanishing magnetic fields, time reversal invariance, 
guarantees that all particle orbits are exactly preserved (except in direction) 
when the injection direction is reversed, irrespective of positioning and
powering errors. Any magnetic field error $\Delta B_x$ 
will move all the beam orbits up or down, depending on the average 
$\langle\Delta B_x\rangle$.
Reversing the beam direction will reverse this shift. The most effective
way of reducing the magnetic field error will be to use the ring as
a ``self-magnetometer'', with the vertical beam position measured by
beam position monitors (BPMs), to adjust magnetic vertical steering 
elements to eliminate the average vertical orbit shift when the beam 
direction is reversed.

Vertical focusing limits the effectiveness of this approach. With no
vertical focusing one could dream that, if the clockwise (CW) beam is
lost in the up-direction then the CCW beam would be lost in the down-direction.
In fact, because of vertical focusing, no beam at all is, in fact, lost. 
The best one can do is to design the ring to maximize the measurable 
vertical orbit shift when the beam direction is reversed.
Nonlinear (octupole) focusing can be strong enough at large amplitudes
to avoid particle loss, yet weak enough at small amplitudes to 
permit precise magnetometry.

The approach to be taken can be motivated by an analogy. The weight of
a fish can be determined by hanging the fish from a spring whose length
is measured by a linear scale. To serve for both small and big fish, with 
scale of convenient length, the spring constant would be strong. For better 
accuracy for small fish, the spring constant would be weak. The ideal spring 
would be weak for small fish but strong for big fish. An octupole
has just this character: its field gradient is negligible for small 
displacements, strong for large displacements.

But accelerator focusing behavior is more complicated. Quadrupoles in 
accelerator lattices, if focusing in one plane, say vertical, are defocusing 
in the other plane. Something like this also applies to octupole focusing,
though, for small amplitudes, a vertically focusing octupole is also
horizontally focusing.
But we need only very weak vertical focusing. We therefore anticipate a
region of phase space near the origin for which the vertical focusing is
weak. Meanwhile the relatively strong horizontal geometric focusing is
much stronger than the octupole focusing. 

\section{Orbit Equations for the Storage Ring Bottle}
\begin{figure}
\centering
\includegraphics[scale=0.5]{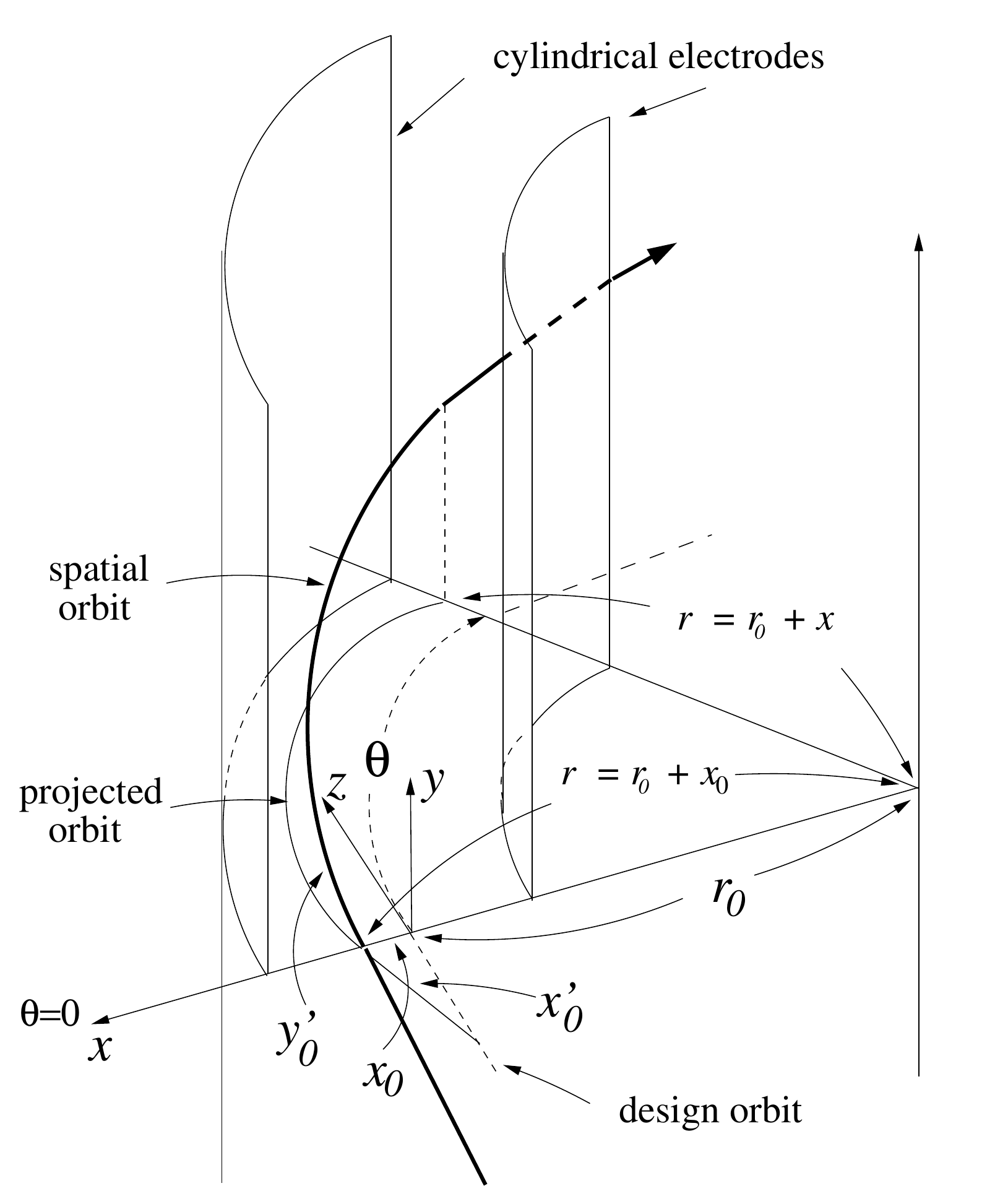}
\caption{\label{fig:CylindricalElectrodes}
The bold curve shows a proton orbit passing through a
curved-planar cylindrical electrostatic bending element. 
The electrode spacing is $g$ and the design orbit
is centered between the electrodes. The electric lattice in
this paper has 281\,m circumference, consisting of 16 indentical 
bends like this, each with arc length 15.7\,m arc length, with
the remainder of the circumference taken up by straight 
sections.}
\end{figure}

The simplest possible electrostatic ring consists of coaxial cylindrical 
electrodes, of inner and outer radii $r_0 \mp g/2$, where $g$ is the gap 
width between the electrodes. A sector of such a ring is shown in
Figure~\ref{fig:CylindricalElectrodes}.

In a cylindrically-symmetric electrostatic potential $\phi(r,z)$,
producing a predominantly-radial electric field $-E{\bf\hat r }$, 
beam particles of charge $e$ move in approximately circular orbits close 
to a horizontal $(x,z)$ plane. The $y$ coordinate is normal to the
plane. On the circular design orbit the centripetal 
electric force is provided by the radial electric field $E_r$;
\begin{equation}
\frac{p_0v_0}{r_0} = -eE_r
\label{eq:Simplest.0}
\end{equation}
With $s$ being a global longitudinal arc 
length parameter, and local Frenet coordinates $(x,y,z)$, 
which are incremental offsets, radial, 
vertical, and longitudinal (i.e. tangential to the design orbit), the 
betatron equations are\cite{ETEAPOT1}
\begin{align}
\frac{d^2 x}{ds^2}
 &=
-\bigg(\frac{1}{r^2_0} + \frac{1}{\gamma_0^2r^2_0}\bigg)\,x
 =
-\frac{1.64}{r^2_0}\,x,
\label{eq:Simplest.1x} \\
\frac{d^2 y}{ds^2}
 &=
0.
\label{eq:Simplest.1y}
\end{align}
Here the numerical value 1.64 has come from choosing $\gamma_0$ appropriate
for frozen spin proton operation (for which
this paper is primarily intended). The relativistic factor 
$\gamma_0=\mathcal{E}/m_pc^2=1/\sqrt{1-v_p^2/c^2}$ has 
the ``magic'' value $\gamma_0=1.25$ at which the proton spins are 
``frozen'' (for example, always parallel to the orbit.) This corresponds 
to proton kinetic energy
$K_p=0.233\,$GeV, velocity $\beta_pc=0.60c$, and momentum 
$p_pc=0.70\,$GeV---relativistic, but not very relativistic. 

Eqs.~(\ref{eq:Simplest.1x}) and (\ref{eq:Simplest.1y})
differ only slightly from the equations for a traditional 
weak focusing magnetic ring. The second shows there is no vertical 
focusing, meaning the beam will eventually
be lost vertically. Vertical stability
can be easily restored by contouring the poles slightly (combined function) 
or introducing vertically focusing quadrupoles (separated function). 
Both of these constitute ``linear focusing'' which, according to
ground rules of the present paper, is not to be allowed (in the interest
of maximal magnetometry sensitivity.) One can object
that the term $\big(x/r^2_0 + x/(\gamma_0^2r^2_0)\big)$ in the $x$ 
equation also constitutes ``linear focusing''. But this is ``geometric''
focusing which (by a convention adopted just for 
this paper) is allowable (in fact, essential). The same terminology
is standard for a cyclotron---it is geometric focusing that 
permits the deviation from the design circular orbit to a slightly
deviant circular orbit to be described as a horizontal betatron 
oscillation crossing 
the design orbit twice per revolution, as if due to focusing.
The extra term $x/(\gamma_0^2r^2_0)$ in the horizontal focusing is 
specific to electrostatic focusing.

Unlike in
a magnetic field, in an electric field the particle velocity is not 
conserved, causing the $\gamma$ factor to deviate from $\gamma_0$ as 
the electric potential deviates from zero. But to a first (good) 
approximation, the orbits are described by Eqs.~(\ref{eq:Simplest.1x})
and (\ref{eq:Simplest.1y}). The only significant effect of the ring 
being electrostatic is that the extra term $x/(\gamma_0^2r^2_0)$ makes 
the horizontal focusing somewhat stronger than in an old-fashioned, 
weak-focusing magnetic ring.

Neglecting the perturbing effects of the short straight sections
between bend elements,
in terms of a ``betatron phase'' $\psi_x(s)$ which, for this simple 
ring, is proportional to $s$, and beta function $\beta_x(s)$, which
is actually independent of $s$, in pseudo-harmonic form, the 
cosine-like $x$ motion for a particle with betatron amplitude 
$a_x$ is
\begin{equation}
x(s)
 = 
a_x\sqrt{\beta_x(s)}\cos\big(\psi_x(s)\big)
 =
a_x\sqrt{\beta_x}\cos\frac{s}{\beta_x}
,
\label{eq:Simplest.2}
\end{equation}
where
\begin{equation}
\psi_x(s)
 = 
\int_0^s \frac{ds'}{\beta_x(s')}
 = \frac{s}{\beta_x},
\label{eq:Simplest.2p}
\end{equation}
Because $\beta_x(s)$ is independent of $s$ the
integral is trivial. In summary:
\begin{align}
x(s)
 &=
A_x\cos
\bigg(
\sqrt{\frac{1 + 1/\gamma_0^2}{r^2_0}}\,s
\bigg),                      \notag \\
y(s)
 &=
A_y + {\rm constant}\,\times s.
\label{eq:Simplest.3}
\end{align}
Matching $x$ arguments gives
\begin{equation}
\beta_x = \frac{r_0}{\sqrt{1+1/\gamma_0^2}},\quad
\Big( 
 \overset{\rm e.g.}{\ =\ }
0.78\times40 =31\,{\rm m}
\Big).
\label{eq:Simplest.4}
\end{equation}
The phase advance per turn $\mu_x$ and the tune 
$Q_x=\mu_x/(2\pi)$ are obtained by
matching $x$ arguments to give
\begin{equation}
Q_x
 =
\frac{r_0}{\beta_x}
 =
\sqrt{1+\frac{1}{\gamma_0^2}}\quad
\Big( 
 \overset{\rm pEDM}{\ =\ }
1.281
\Big),
\label{eq:Simplest.5}
\end{equation}
and $Q_y=0$.

\subsection{Distributed Octupole Field}
One way or another, the vertical motion has to be stabilized.
To do this we introduce an octupole field, assumed to 
be uniformly distributed around the ring, with local octupole 
moment per unit length $o=O/(2\pi r_0)$. (For numerical simulations
shown later in this paper the ``uniform octupole distribution'' 
is modeled as 16 lumped octupoles, one in each straight section.)
The orbit equations become
\begin{align}
\frac{d^2x}{ds^2}
 &= 
-\frac{1.64}{r^2_0}\,x
-\frac{o}{6}\,(x^3-3xy^2) - qx + f_x,
\label{eq:Octupole.1x} \\
\frac{d^2y}{ds^2}
 &= 
 \frac{o}{6}\,(3x^2y-y^3) + qy + f_y.
\label{eq:Octupole.1y}
\end{align}
As well as the octupole terms\cite{USPASmanual}, for later convenience,
we have included other small terms that can be neglected 
initially. A possible distributed ``trim'' quadrupole is represented 
by the terms $-qx$ and $+qy$; positive $q$-value introduces
(very weak) horizontal focusing and vertical defocusing.
Also added are constant ``force'' terms $f_x$ and $f_y$. 

This paper concentrates on the added term $f_y$ which represents
an unknown magnetic ``force'' that needs to be cancelled. 
The term $f_x$ will eventually be set to zero based on the 
following comments. Though $x$ and $y$ are locally orthogonal, 
only $y$ is fixed globally, while $x$ is a Frenet coordinate, 
always radial. A constant radial force is best treated as a 
change in radius of curvature, leaving the orbit plane 
invariant, but changing the design orbit radius. A constant vertical 
force, on the other hand, tends 
to make an otherwise-planar orbit helical, leading to a 
gradual vertical displacement of the whole orbit.

The octupole terms, as well as being nonlinear, 
also couple the $x$ and $y$ motion. Solving these equations 
analytically and in general for all amplitudes is
clearly impossible, even if for no other reason than that the motion
will become chaotic for sufficiently large amplitudes. If the
equations are to describe anything practical, they will have to
apply only to appropriately small amplitudes. Fortunately this
will be applicable for electrostatic rings appropriate for 
measuring EDMs. To obtain large electic field $E_y$ it is important
for electrode gap $g$ to be small. This forces the horizontal
ring acceptance to be small. On the other hand it is feasible
for the vertical acceptance to be almost arbitrarily large. 
Unbalanced acceptance like this is favorable for octupole focusing. 

\subsection{Guiding Center Approximation}
We are interested primarily in two effects influencing the vertical orbit 
motion. One is the role played by the electrostatic octupole field 
present for limiting the vertical motion. Each particle gyrates
rapidly while advancing slowly in the vertical direction, 
oscillating up and down as the
octupole field occasionally causes it to reflect, down from 
the top or up from the bottom. The other effect to be included
is the systematic vertical orbit shift caused by any residual
constant radial magnetic field. Sensitivity to this shift, opposite 
for CW and CCW orbit circulation
direction, is the basis of the self-magnetometry function.

Language here has been chosen to suggest 
the non-relativistic orbit of a particle gyrating around a field line
in a slowly varying magnetic field,
for example in a magnetic bottle or in the earth's ionosphere. With each 
orbit gyrating rapidly around a ``guiding center'', it is
adequate to describe the motion of the guiding center, rather
than following the charged particle itself. In our relativistic
storage ring with ultraweak vertical focusing, each particle
moves in a helical orbit, always in an almost closed circle of 
radius $r_0$, perhaps 40\,m. The guiding center hardly moves from 
the center of the ring, mainly just up and down by, at most, a few
centimeters.

The guiding center formalism is appropriate
for describing a storage ring relativistic bottle with uniform vertical
magnetic field. This theory is sketched concisely in the 
Section~\ref{sect:Appendix} appendix. To corroborate applying this theory 
to magnetic storage rings some TEAPOT\cite{TEAPOT}\cite{UAL} 
simulations of a 
magnetic storage ring bottle are shown later in this paper. 
But we are primarily interested in an the electric storage ring 
bottle, to be modeled later using ETEAPOT\cite{ETEAPOT1}. First 
the equations are solved approximately.

\subsection{Fast-Slow Approximation\label{sect:FastSlow}}
Ordinarily, in storage rings, a fast-slow approximation is used in
which synchrotron (energy) oscillations are ``slow'' enough to be
treated as ``adiabatic'' (i.e. essentially constant) as regards 
their influence on $x$ and $y$ betatron oscillations. Slightly
different approximations are appropriate in the present
situation. Only $x$ oscillations are fast; longitudinal $z$ and 
vertical $y$ oscillations are slow.

Part of the justification for this picture is based on the 
particular configuration being discussed, which can be refreshed 
by reviewing Figure~\ref{fig:CylindricalElectrodes}. Because 
practical electric forces are vastly weaker than magnetic forces,
to produce sufficiently large electric field
the electrode gap $g$ is necessarily small, 2 or 3\,cm. This 
restricts the horizontal amplitude $a_x\sqrt{\beta_x}$ to be quite 
small. On the other hand, for any stable vertical 
oscillations, the absence of linear focusing will
certainly cause the vertical tune $Q_y$ to be very small.
Furthermore, since the electrodes can be almost arbitrarily high, 
the vertical coordinate $y$ is allowed to be quite large,
much larger than $x$. 

This suggests approximating $x^2$ by its average value while 
treating the $y$ motion.
In Eq.~(\ref{eq:Octupole.1y}) we set $x^2=\sigma_x^2$,
where $\sigma_x\approx g/4$ is an approximate r.m.s. 
value of $x$ for a parabolic-shaped proton beam in a chamber with 
width $g$. The result is 
\begin{equation}
\frac{d^2y}{ds^2}
 = 
\bigg(\frac{og^2}{32} + q \bigg)\,y + \frac{o}{6}\,y^3 + f_y.
\label{eq:Octupole.2}
\end{equation}
(The ``trim quadrupole'' term $+qy$ reserves the possibility of 
empirically cancelling the $y$-term, at least approximately, by
tuning the value of the parenthesized term to zero.) 

\begin{figure}[h]
\centering
\includegraphics[scale=0.33, angle=-90]{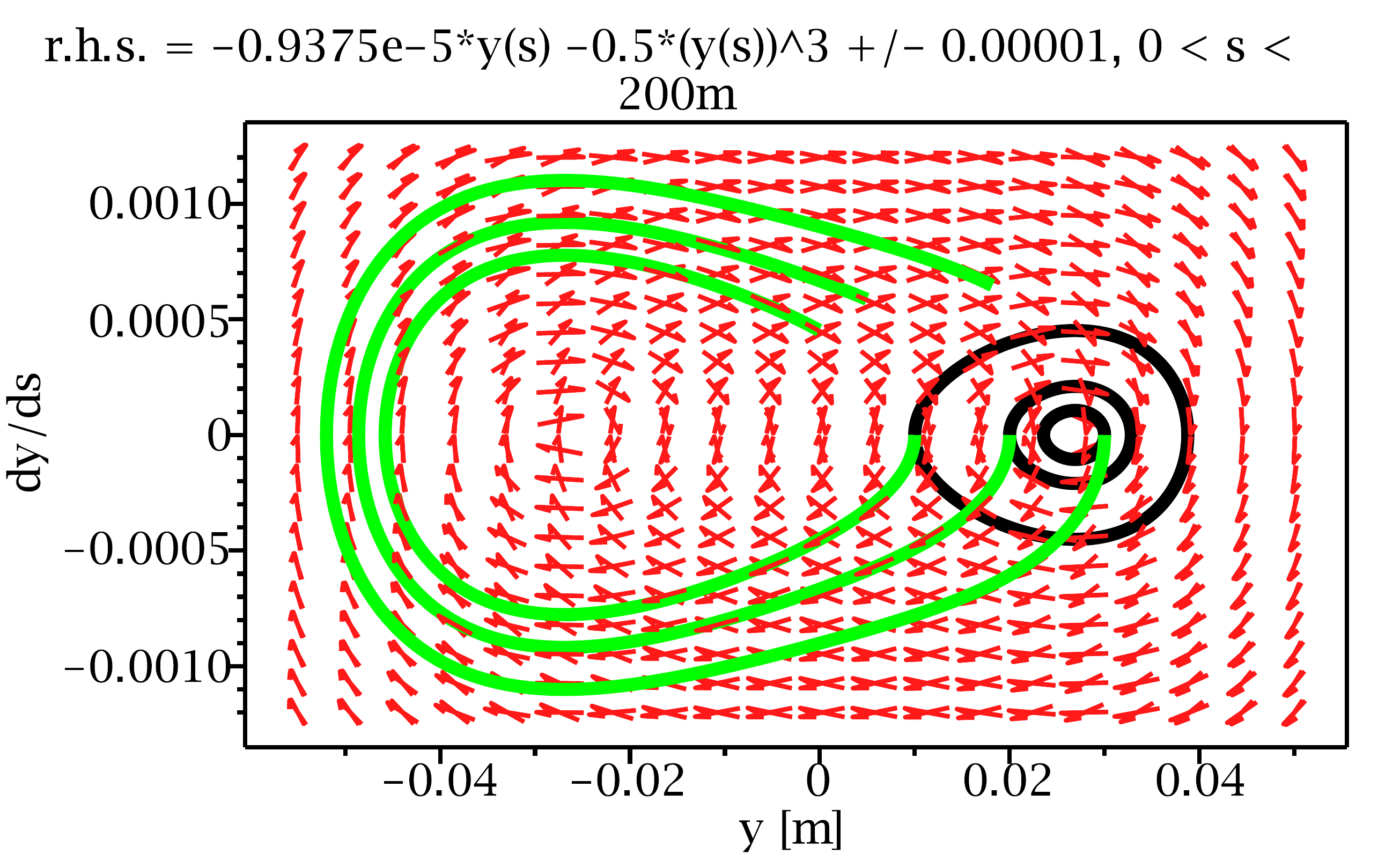}
\caption{\label{fig:B_rSensitivity_p}Vertical phase space trajectory
dependence on constant vertical force in storage ring bottle with
radial geometric focusing plus octupole (but no quadrupole) focusing. 
Curves with +/- constant force are 
black/green. ``r.h.s.'' stands for the right hand side of Eq.~(\ref{eq:Octupole.2}). 
The magnetometer sensitivity is proportional to the displacement of black 
relative to green. The plot was obtained using MAPLE.
}
\end{figure}

Though now decoupled, this equation 
is still nonlinear. However, since the right hand side is a function 
only of $y$, the orbit description can be simplified by 
introducing (conserved) ``energy'' $h$,
\begin{equation}
h
 = 
\frac{1}{2}\,
\bigg(\frac{dy}{ds}\bigg)^2 - \bigg(\frac{og^2}{64} + \frac{q}{2}\bigg)\,y^2
        - \frac{o}{24}\,y^4 - f_yy.
\label{eq:Octupole.3} 
\end{equation}
(The energy symbol ``h'' introduced here has been chosen the same as
an analogous symbol for energy, in the Section~\ref{sect:Appendix} 
appendix for mnemonic convenience. Though the ``h'' roles are analogous in 
the two contexts, their physical dimensions differ for inessential 
reasons of convenience. A dimensional factor needed to restore 
dimensional consistency and a relativistic correction, not very
important for $\gamma=1.25$ protons, will be introduced later.)
Much of the vertical motion can then be inferred from curves of
constant $h$ in $(y,dy/ds)$ phase space. Numerical solutions,
using MAPLE, are shown in Figure~\ref{fig:B_rSensitivity_p}.
As well as exhibiting three amplitudes, the figure also allows for
the presence of a small $f_y$ term, of either sign, black for
positive, green for negative. The ``vertical force sensitivity''
is the beam centroid deviation divided by the vertical ``force''
which, in this case is 
$0.06/0.00001=0.6\times10^4$ (in the units 
implied by the numerical form given at the top of the figure).

\begin{figure}[h]
\centering
\includegraphics[scale=0.33, angle=-90]{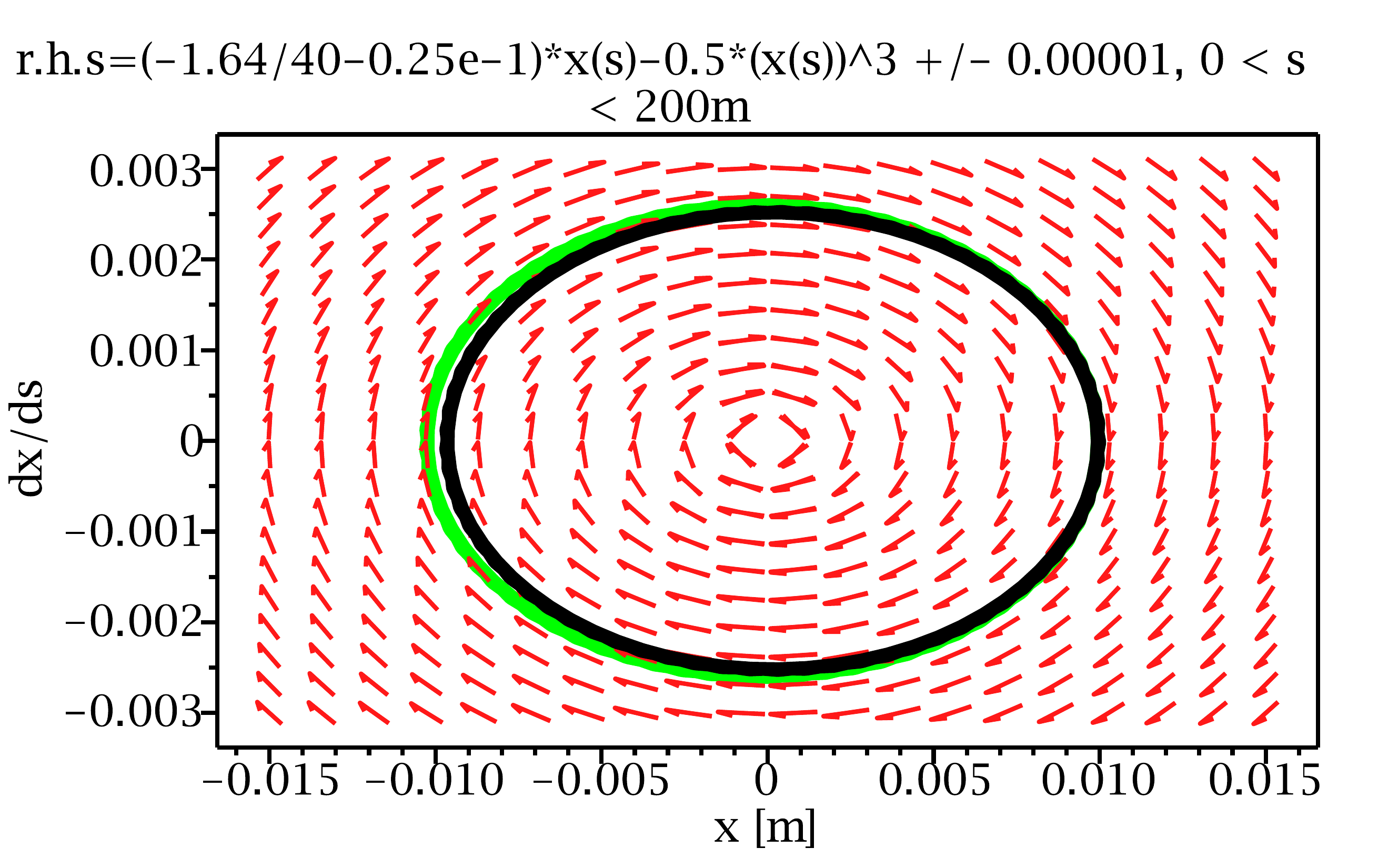}
\caption{\label{fig:B_rLinearSensitivity}Sensitivity of horizontal 
phase space trajectory to constant horizontal force in storage ring bottle 
with radial ``geometric'' focusing plus octupole focusing.
``r.h.s.'' stands for the right hand side of Eq.~(\ref{eq:Octupole.4}).
Curves with +/- constant force are black/green. 
The magnetometer sensitivity is proportional to the displacement of 
black relative to green. The plot was obtained using MAPLE.}
\end{figure}

The corresponding $x$-equation, with $y=y_c$ (with $y_c$ treated
as varying adiabatically, and hence effectively constant, with
value $0$) is
\begin{align}
\frac{d^2x}{ds^2}
 &= 
-\bigg(\frac{1.64}{r^2_0} + \frac{o}{6}\,y_c + q\bigg)\,x 
  - \frac{o}{6}\,x^3 + f_x.
\label{eq:Octupole.4}
\end{align}
Numerical determinations of the phase trajectories are
shown in Figure~\ref{fig:B_rLinearSensitivity}. 

Comparing Figures~\ref{fig:B_rSensitivity_p} and 
\ref{fig:B_rLinearSensitivity}, one sees that the vertical
sensitivity to $f_y$ is two orders of magnitude greater than
the horizontal sensitivity to $f_x$. Of course this is
because the horizontal focusing is of linear order, while
the vertical focusing is of octupole order. 
In ``typical'' storage rings the horizontal and vertical 
focusing strengths
are comparable, which is manifested by their radial and 
vertical tunes $Q_x$ and $Q_y$ 
being comparable in magnitude. Our storage ring bottle 
lattice has therefore achieved the goal of increasing the 
vertical force sensitivity by a factor of about 100,
compared to typical rings. 

This comparison is somewhat misleading howerer, for reasons 
alluded to earlier. The main effect of reversing $f_x$ suddenly 
is a change in gross orbit radius. But there is also the net 
radial orbit shift shown in Figure~\ref{fig:B_rLinearSensitivity}.
The global direction of this shift would depend on the particle 
azimuthal position when the force was reversed. As such
the shift is not directly commensurate with the vertical
magnetometer orbit shift which is the subject of this paper.
Also this comparison does not take advantage of potentially
increased magnetometer sensitivity using quadrupole $q$
to tune the linear focusing coefficient more nearly to zero.


\subsection{Octupole Restrained Motion}
Relativistic orbits in a magnetic bottle are analysed in the appendix,
following formalism standard in plasma physics. However, 
the adiabatic invariance of the product of
effective current $\times$ enclosed area, which is central to the 
magnetic bottle treatment, may not be valid for an electric bottle, 
especially because the electric potential depends
on position. Nevertheless, since the orbits have already been seen to
be so similar, one expects reflections down from the top and up 
from the bottom to be subject to similar analyses.

When it passes through the $y=0$ horizontal design plane, a given 
particle has vertical momentum $p_{y0}$ which is, let us say, positive,
and certainly very small compared to its total momentum $p_0$. Because
$p_{y0}<<p_0$, the helical orbit is scarcely distinguishable from a perfect
circle. For many turns, because the octupole is so weak near the origin, 
$p_y$ remains essentially constant. But, eventually, the integrated 
effect of the octupole field is sufficient to reflect the vertical 
motion. The magnitude $y_{\rm t.p.}(p_{y0})$ of the vertical turning point 
of the $y$-motion clearly increases with increasing $|p_{y0}|$.

Comparing Eq.~(\ref{eq:Octupole.3}) with Eq.~(\ref{eq:MagTrap.12}) in the
Section~\ref{sect:Appendix} appendix, one sees that vertical motion
(perpendicular to electric field) in an electric bottle, is subject
to much the same analysis as longitudinal motion (parallel to magnetic
field lines) in a magnetic bottle. In each case the particle moves
more or less freely parallel to the axis of the bottle, but is
occasionally restrained by reflection from turning points at 
one or the other end 
of the bottle. In the electric bottle the ``vertical energy'' of a
particle with slope $y'_0$ at the center of the bottle is 
$h_0=\gamma_0{y_0'}^2/2$; use of the non-relativistic energy formula 
(slightly modified by the $\gamma_0$ factor accounting for inertial
mass) will be justified shortly. At the turning points, 
at the bottle ends, $y'=0$.
Substituting into Eq.~(\ref{eq:Octupole.3}), with $q=0$ and $f_y=0$ 
produces
\begin{equation}
y_{tp}^4 + \frac{3}{8}\,g^2y_{tp}^2 + \frac{12}{o}\,\gamma_0{y_0'}^2 = 0.
\label{eq:Reflection.1} 
\end{equation}
(For magic energy protons the relativistic correction is small, and
is being neglected in this paper. But,
for magic energy electrons the relativistic factor is $\gamma_0=30$, 
giving a significant relativistic correction even if the
vertical component of velocity is non-relativistic.)
Solving for $y_{tp}$ produces 
\begin{equation}
y_{tp}
 =
\frac{\pm 1}{\sqrt{2}}\,
\sqrt{
-\frac{3}{8}\,g^2 + \sqrt{\frac{9}{64}\,g^4-\frac{48}{o}\,\gamma_0{y_0'}^2}}, 
\label{eq:Reflection.2} 
\end{equation}
or, treating $g$ as negligibly small, $o\approx-12\gamma_0(y'_0/y_{\rm tp}^2)^2$.
For real-valued turning points the octupole strength coefficient $o$
certainly has to be negative.

This analysis of a relativistic electric bottle has relied on 
artificially suppressing coupling between $x$ and $y$ motions.
This makes the analysis significantly less robust than the treatment
of non-relativistic magnetic traps in the appendix; that treatment
utilizes adiabatic invariance to incorporate transverse
motion into the longitudinal description. The
adiabatic invariant treatment accurately performs the averaging brought 
about by the gyrating motion around field lines. Though less elegant,
our analysis has assumed (on intuitive grounds) that a more or less 
equivalent averaging is accurately represented by the replacement 
$x^2\rightarrow\langle x^2\rangle$ in Eq.~(\ref{eq:Octupole.1y}).  
Numerical simulations described in following sections 
seem to confirm the essential validity of this assumption.

Though the analogy between our electric bottle and the 
plasma physics magnetic bottle analysed in the appendix is close, it is 
not perfect. If the lengths of the both bottles are defined to be $2y_{tp}$, 
then the amplitude of the motion within
the electric bottle increases with increasing particle slope $y_0'$, 
up to a limit beyond which the particle escapes the trap longitudinally. 
That is, large amplitude particles are lost out the ends of 
the relativistic electric 
trap. It is low amplitude particles which, because of their too low
orbital gyration magnetic moment, that can leak out from the ends of a 
non-relativistic magnetic plasma trap.

In a subsequent section of this paper the relativistic electric trap is 
simulated using ETEAPOT and, following that, the relativistic magnetic 
trap is simulated using TEAPOT numerical simulation. The qualitative 
behaviors of these relativistic traps are essentially the same.

But it is only the electric trap for which
there is an obvious and unique application. This application, of course,
is the proton (or electron) EDM measurement. The high sensitivity
to radial magnetic field $B_r$, along with the opposite vertical 
displacement that $B_r$ causes for CW and CCW beams, enables the
average value $\langle B_r\rangle$ to be cancelled to high precision.

\section{Self-Magnetometer Precision\label{sect:Magnetometer}}
\subsection{Compensation Procedures}
The need for self-magnetometry comes from the requirement to
minimize the average radial magnetic field $\langle B_r\rangle$ 
in an electric bottle measurement of the EDM of the proton (or the 
electron).  $\langle B_r\rangle$ is to be obtained by  
measuring the average vertical orbit displacement that results
when the beam direction is switched from clockwise to 
counter-clockwise. Though EDM run durations will be of order
1000\,s, the $\langle B_r\rangle$ compensation can probably be 
completed in a measurement sequence taking several seconds.

We assume there is a compensation coil whose integrated value
$\langle B_r^{\rm comp}\rangle$ cancels $\langle\Delta B_r\rangle$ 
to the same precision that it can be 
measured\footnote{Exactly how the $B_r$ magnetic field compensation
coils are to be designed remains a serious issue.
Almost certainly all elements have to be wired in series, powered 
by a single DC current of, perhaps, 10\,mA.
The reason for this restriction 
is that the $\langle B_r^{\rm comp}\rangle$ 
compensation field has
to be reliably reversible, within about one second, with accuracy 
better that one part in $10^9$, with correspondingly stationary
mechanical stability, as part of the EDM measurement.
Ideally the circuit would consist of two current loops around
the ring, one above, one below the beam. With AC powering
at frequency low enough for vertical orbits to follow, this
circuit will provide an accurate magnetometer calibration.
AC-coupled (to guarantee reversal symmetry) at, say 0.001\,Hz
frequency, this process might even be the basis for the actual EDM
measurement.}. 
To produce a numerical 
result we will assume the all-electric proton lattice whose behavior 
is simulated in later Section~\ref{sect:ElectricBottle}.

The basic magnetometry idea is that the unknown magnetic field
$\langle\Delta B_r\rangle$, though insignificant over a single turn,
acts monotonically over multiple turns, constantly tending to change 
the vertical orbit slope. Lattice element forces, also small near 
the origin, but producing net focusing, are just 
large enough to prevent the eventual loss of injected particles. 
It is the displacement of the centroid of the injected bunch that 
provides the measurement of $\langle\Delta B_r\rangle$. 

One thing that makes the process difficult
to analyse is that, even if the injected bunch is ideally well
collimated it is probably mismatched, and quickly filaments 
to a broader equilibrium betatron distribution. During this
evolution the particles are subject to oscillatory restoring 
forces large compared to the magnetic force being measured. 
The BPM electronics will be required to measure a vertical 
centroid shift very small compared to the r.m.s. bunch height 
$\sigma_y$.

Another complication, neglected so far, is that there
will surely also be a systematic non-zero vertical electric field
error $\langle\Delta E_y\rangle$ that will, itself, produce a vertical
shift of the beam centroid, probably large compared to the magnetic
shift we have been discussing---for example because of imperfect 
electrode shapes or positioning.
This shift will be nulled by $\langle\Delta E_y^{\rm comp}\rangle$ 
compensation provided by multiple vertically-separated electrodes, 
distributed as finely and uniformly as possible, in straight sections 
around the ring. This compensation will be paired with the
$\langle\Delta B_r^{\rm comp}\rangle$ magnetic compensation, also
distributed as uniformly as possible around the 
ring.

This pairing is 
natural since both of these field components deflect the beam vertically.  
The result of perfect empirical adjustment would be
\begin{align}
\langle B_r\rangle 
 &=
\langle\Delta B_r\rangle 
 +
\langle B_r^{\rm comp}\rangle = 0 \label{eq:Compensate.B_r}  \\
\langle E_y\rangle 
 &=
\langle\Delta E_y\rangle
 +
\langle E_y^{\rm comp}\rangle = 0.
\label{eq:Compensate.E_y}
\end{align}
For single beam operation, what has so far been referred to as a 
``magnetometer'' actually responds to the sum of electric
and magnetic vertical forces. The actual magnetometry
functionality only comes from measuring the difference between
CW and CCW beams.

It has to be recognized then, that
when a single beam centroid position is adjusted, with either 
$\langle E^{\rm comp}_y\rangle$ or 
$\langle B^{\rm comp}_r\rangle$, 
it is actually the sum of the electric and magnetic vertical 
forces that is being cancelled. 

During each run the vertical electric field average
$\langle\Delta E_y\rangle$ is likely to drift enough to require
feedback stabilization. Certainly any such feedback has to rely
on electric (not magnetic) vertical steering. 
It is important to realize that, by itself, the presence
of a $\langle\Delta E_y\rangle$ field error does not
affect the EDM measurement---any precession
this field causes is around a vertical axis and will be 
cancelled as part of a phase-locked polarization control 
feedback systems. In this sense the $\langle E^{\rm comp}_y\rangle$
value is unimportant.

For the EDM measurement the $\langle B^{\rm comp}_r\rangle$ 
compensation coil will be adjusted to null the
vertical separation of CW and CCW circulating beams. 
One candidate for proton EDM 
measurement\cite{BNLproposal}\cite{ETEAPOT2} employs simultaneously 
counter-circulating proton beams for, perhaps, $10^3$ seconds, 
or $10^9$ turns. Both the (non-critical) electric 
and the (critical) magnetic nulling could then be accomplished 
in real time during a single fill. This will be ideal for 
self-magnetometry, which should work splendidly. Furthermore 
the method will work well even with unequal beam currents.

Unfortunately, simultaneously counter-circulating beams bring 
in all manner of other experimental complications. Perhaps the most
troublesome is that Wien filters (needed for some operations)
are directional. Adjusted to have no effect on orbits of one beam, 
they deflect counter-circulating orbits. 
Also, simultaneously counter-circulating beams cannot be used to 
measure the EDM of particles other than proton and electron, 
since both electric and bending is required in the ring to
freeze their spins. Nevertheless, the
single beam cancellation of radial magnetic field will be
satisfactory.

The alternative to simultaneously counter-circulating beams is
nulling $\langle\Delta B_r\rangle$ between every data collection 
run. This would employ a sequence of brief runs alternating 
between CW and CCW circulation, with the nulling improved 
iteratively. This nulling will be performed using
$\langle B^{\rm comp}_r\rangle$ compensation. 

Wien filter directionality also complicates this magnetic
field nulling. Wien filters would probably be turned
off during $\langle B_r\rangle$ cancellation.
Then spurious precession due to the Wien filter itself
would need to be accounted for separately, which is
at least a nuisance. The real problem, though, comes from 
unknown magnetic field changes occurring during actual  
EDM runs and interpretable as being due to EDMs. 
Neverthless, alternating between CW and CCW runs 
separated by brief magnetic field compensation is
calculated\cite{FreqDomainEDM} to produce very 
accurate EDM measurements.

\subsection{Calculated Self-Magnetometer Precision}
In Eq.~(\ref{eq:Octupole.2}), as is customary in lattice theory, 
the ring dynamics have been
expressed in purely geometric terms, with no reference whatsoever
to absolute momenta, nor electric nor magnetic field strengths. 
To estimate magnetometer precision, the first task is to establish 
these quantities for practical application. Parameters 
to be evaluated in this section are recorded in 
Table~\ref{tbl:BottleParameters}.
\begin{table}[h] 
\begin{tabular}{|c|c|c|c|c|c|}  \hline
parameter            &  symbol       &    unit   &    value   \\ \hline
ring radius          &  $r_0$        &     m     &    40      \\
electrode gap        &  $g$          &     m     &   0.03     \\ \hline
nom. initial slope   &  $y'_0\equiv dy/ds|_0$ &  &   0.001    \\
nom. turning point   &  $y_{\rm t.p.}$ &     m     &   0.02     \\ \hline
octupole coefficient &  $o$          & m${}^{-4}$ &  -0.003    \\ 
trim quadrupole      &  $q$          & m${}^{-2}$ &   0.005    \\ \hline
constant coefficient &  $k_0$        &    m       & $4\pi^2r_0\beta_0(c\Delta B_r/E_0)$ \\
                     &               &            & $2.8\times10^4\Delta B_r[{\rm T}]$  \\ \hline
linear coefficient   &  $k_1$        &            & -0.00033  \\
cubic coefficient    &  $k_3$        & m${}^{-2}$  & -31.6     \\ 
 \hline
\end{tabular}
\medskip
\caption{\label{tbl:BottleParameters}Typical parameters for relativistic
electric storage ring bottle. A numerical value for the magnetic error
that would roughly mimic the effect of an EDM equal to $10^{-29}$\,e-cm gives
$k_0=k_0^{\rm nom}\approx0.5\times10^{-12}$\,m.}
\end{table}
In terms of the vertical angle $\theta_y=dy/ds$ (paraxial approximation)
that a particle makes 
with the horizontal plane, Eq.~(\ref{eq:Octupole.2}) can be expressed as 
\begin{equation}
\frac{d\theta_y}{ds}
 = 
 \frac{og^2}{32}\,y + \frac{o}{6}\,y^3 + f_y,
\quad\hbox{where}\quad
\theta_y\equiv y'\equiv \frac{dy}{ds}.
\label{eq:Magnetometer.1}
\end{equation}
Here $f_y$ can be interpreted as the vertical angular change,
per meter of particle path length, caused by a constant vertical
force. We assume $f_y$ is caused by an (unknown) radial magnetic field 
$\Delta B_r$, which we take to be uniformly distributed around a 
circular ring of radius $r_0$.

For numerical convenience the independent variable will be changed
to turn number $n$, defined by $s\rightarrow 2\pi r_0 n$. 
Eq.~(\ref{eq:Octupole.2}) is transformed to 
\begin{equation}
\frac{d^2 y}{dn^2}
 \equiv
\frac{dy'}{dn}
 = 
k_0 + k_1 y + k_3 y^3,
\label{eq:Magnetometer.1p}
\end{equation}
where, now, $y'\equiv dy/dn$
has dimensions of length, and is the instantaneous 
vertical position advance per complete turn. 

\bigskip
\noindent{\bf Digression: why quadrature solution is impractical.\ }
Direct solution of the differential equation could proceed as 
follows. Setting $y(0)=0$,
Eq.~(\ref{eq:Magnetometer.1p}) can be solved for $dy/dn$ as a 
function of $y$; then $dn/dy(y)=1/(dy/dn)$;
\begin{equation}
\frac{dn}{dy}\,(y)
 = 
\frac{2}{\sqrt{8 k_0 y + 4 k_1 y^2 + 2 k_3 y^4 + 4\big(dy/dn|_0\big)^2}}.
\label{eq:Magnetometer.1q}
\end{equation}
This can be integrated to obtain $n(y)$;
\begin{equation}
n(y)
 = 
\int_0^y
\frac{2 dy'}
{\sqrt{8 k_0 y' + 4 k_1 {y'}^2 + 2 k_3{y'}^4 + 4\big(dy/dn|_0\big)^2}}.
\label{eq:Magnetometer.1r}
\end{equation}
Here, third meaning so far, $y'$ is dummy integration variable.
The integration can be performed analytically and
after inversion, this yields $y(n)$ in closed form.

This approach to solving the equation of motion is valid for short time
intervals but oscillatory forces make the approach impractical for
large values of $n$. The numerical value of $k_0$ will be extremely 
small. Its presence can only be felt after a large number of turns.
For example, if $k_0=10^{-14}$\,m were the only ``force'',
then $y=k_0n^2/2$ solves Eq.~(\ref{eq:Magnetometer.1p}).
This suggests an accumulated displacement of 0.005\,m 
after $10^6$ turns. Regrettably, this greatly exaggerates the
actual shift. There can be no accumulation proportional 
to $n$, much less $n^2$. The restoring forces cause oscillatory 
motion that limits any accumulation due to a constant force. 

Adversely, it can be noted that a vertical centroid injection 
``slope'' error of $5\times10^{-9}$\,m (orders of magnitude smaller 
than could be practically achieved) would, in the absence of vertical 
focusing, give the same  0.005\,m accumulated displacement over 
the same $10^6$ turns. The inherent oscillatory motion caused
by focusing terms makes this estimate similarly misleading.
\noindent{\bf End of digression. }
\bigskip

In ordinary simple harmonic motion $k_3=0$ and the
condition for vanishing ``acceleration'' produces 
$y_{\rm f.p.}$=$-k_0/k_1$ as the ``fixed point''.
Magnetometer sensitivity is increased by decreasing 
the linear focusing coefficient $k_1$ to increase
$y_{\rm f.p.}$. The whole point 
of octupole focusing is to carry this to the extreme. 
Significant simplification results if $k_1=0$. 
(It is the purpose for ``trim quadrupole'' $q$.)
Then the fixed point condition is
\begin{equation}
y_{\rm f.p.} = -\bigg(\frac{k_0}{k_3}\bigg)^{1/3}.
\label{eq:Magnetometer.1pp}
\end{equation}
Because the ``force'' is $k_3y^3$ the octupole
focusing effect can be large for large $y$ amplitude
even for small $k_3$. In this way the magnetometer can
be highly sensitive while preserving adequate focusing 
for large amplitude (where it is needed to reflect
the orbit).

(Incidentally, it can be noted that multipole elements of even
higher order than octupole would be subject to identical analysis.
The only requirement is that, like octupole, the
multipole order has to be a multiple of 8 to have small amplitude
focusing in both horizontal and vertical planes. 
With increasing multipole order the region of nearly vanishing 
focusing near the origin 
becomes increasingly large. In this way the magnetometer 
sensitiviy could be made almost arbitrarily high, but other problems 
would be likely. Best of all, but not physically possible,
would be perfectly reflective walls top and bottom.)

The coefficients $k_0$, $k_1$, 
and $k_3$ are to be determined next.
The vertical equation of motion,
for a particle of charge $e$, velocity $v_0$, momentum $p_0$, is 
\begin{equation}
\frac{dp_y}{dt}
 =
ev_0\Delta B_r.
\label{eq:Magnetometer.2}
\end{equation}
The vertical angular increment per revolution 
(of time duration $2\pi r_0/v_0$) is
\begin{equation}
\Delta\theta_{y,1}
 =
\frac{\Delta p_{y,1}}{p_0}
 =
\frac{2\pi r_0(c\Delta B_r)}{p_0c/e},
\label{eq:Magnetometer.3}
\end{equation}
where the ``1'' subscript indicates one turn, and the
factors have been grouped for convenience with units.
The main storage ring bending force is provided by electric 
field $E_0$, in terms of which the total particle momentum can 
be expressed as
\begin{equation}
p_0c/e = \frac{r_0E_0}{v_0/c}.
\label{eq:Magnetometer.4}
\end{equation}
We then obtain, expressed as vertical displacement advance 
per turn,
\begin{equation}
k_0
 =
2\pi r_0\Delta\theta_{y,1}
 =
4\pi^2 r_0\beta_0\Big(\frac{\Delta B_r}{E_0/c}\Big).
\label{eq:Magnetometer.5}
\end{equation}
Expressing $k_0$ in this form is convenient because the
denominator factor $E_0/c$ acting on the nominal EDM
can be visualized as producing the intended EDM precession 
and the numerator factor $\Delta B_r$ acting on the
proton MDM can be visualized as producing the spurious
precession. The relative torque of
magnetic field $B$ acting on the proton MDM and electric 
field product $cE$ acting on nominal EDM is
$\eta_{\rm EM}^{(p)}=0.53\times10^{-15}$. So a nominal
value $k_0^{\rm nom}$ that mimics the full nominal EDM
signal is obtained as 
\begin{equation}
k_0^{\rm nom}
 =
4\pi^2 r_0\beta_0\times 0.53\times10^{-15}
\ 
\Big(\overset{\rm e.g.}{\ =\ } 0.5\times10^{-12}\,{\rm m}\Big).
\label{eq:Magnetometer.5p}
\end{equation}
The other coefficients in Eq.~(\ref{eq:Magnetometer.1p})
are given in Table~\ref{tbl:BottleParameters}. They
depend primarily on the octupole
strength parameter $o$, which has to be chosen to give
satisfactory bottle performance. Its listed value is appropriate 
for a particle with a typically large initial vertical slope
such as $dy/ds|_0\overset{\rm e.g.}{\ =\ }0.001$, 
starting at $y_0=0$, to be ``turned'' at a convenient
turning point such as
$y_{\rm t.p.}\overset{\rm e.g.}{\ =\ }0.02\,$m. This initial
slope corresponds to 2-sigma, for a vertical emittance
$\epsilon_y=10^{-5}$\,m, assuming (Twiss parameter) 
$\beta=r_0=40$\,m. Later the value of $o$ can be 
refined---smaller for better magnetometer sensitivity,
larger for larger angular acceptance.

According to Eq.~(\ref{eq:Reflection.2}), setting $g=0$ for
simplicity, the octupole strength can be estimated in terms of 
the turning point coordinate by 
\begin{equation}
o = -12\gamma_0\,\Big(\frac{y'_0}{y^2_{\rm t.p.}}\Big)^2.
\label{eq:Magnetometer.6}
\end{equation}
(The trim quadrupole strength $q$ can be adjusted to remove the linear
focusing term in Eq.~(\ref{eq:Octupole.2}), which is equivalent to
setting $g=0$ and $k_1=0$. This amounts to tuning the vertical linear
focusing exactly to zero in order to maximize the magnetometry sensitivity.
This is operationally hazardous however since, for a particle with
tiny horizontal amplitude and (temporarily) tiny vertical amplitude,
the vertical amplitude (temporarily) grows exponentially until
the octupole focusing kicks in. The phase space region where this
potentially chaotic behavior is possible clearly has to be kept small.
In numerical investigation, tiny figure-eight phase space trajectories 
appear near the origin the origin.)

Eq.~(\ref{eq:Magnetometer.6})  determines octupole coefficient $o$ and, 
from it, the remaining entries
in Table~\ref{tbl:BottleParameters}. One can check, at $y_{\rm t.p.}$, that
the $k_3y_{\rm t.p.}^3$ octupole restoring force is much greater than the
$k_1y_{\rm t.p.}$ quadrupole restoring force, which validates having neglected
$g$ in Eq.~(\ref{eq:Reflection.2}). 

\begin{figure}[h]
\centering
\includegraphics[scale=0.33, angle=-90]{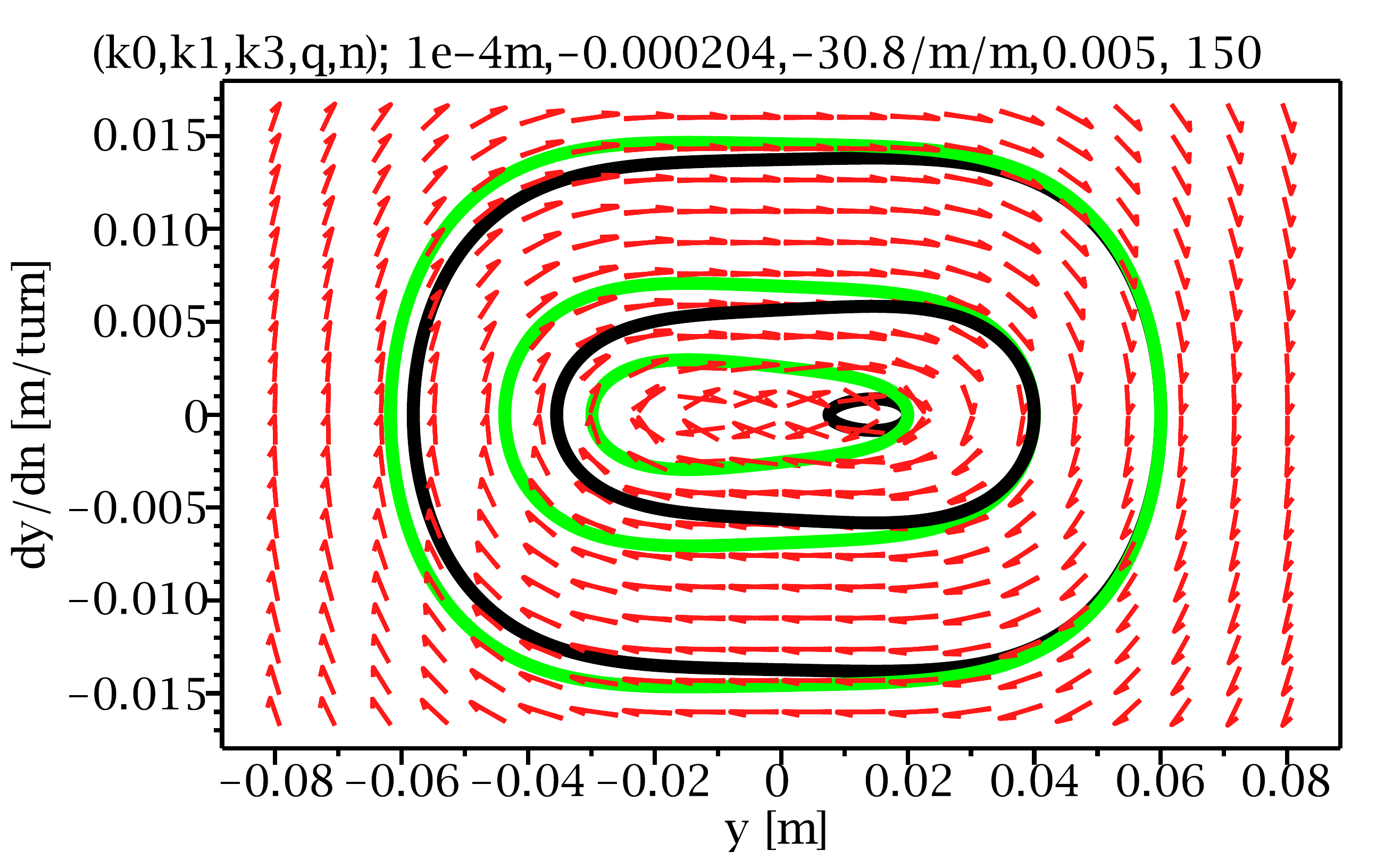}
\includegraphics[scale=0.33, angle=-90]{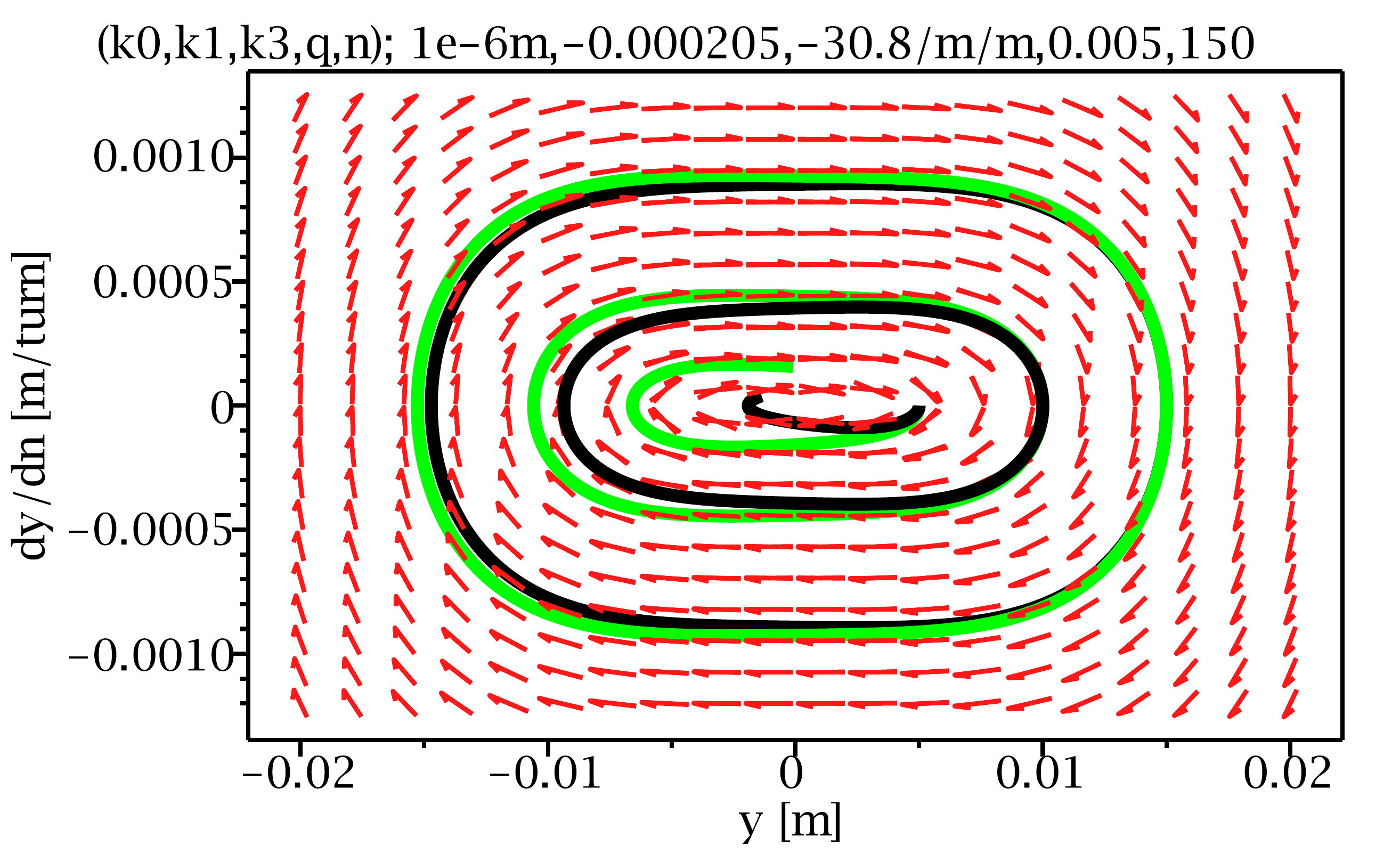}
\includegraphics[scale=0.33, angle=-90]{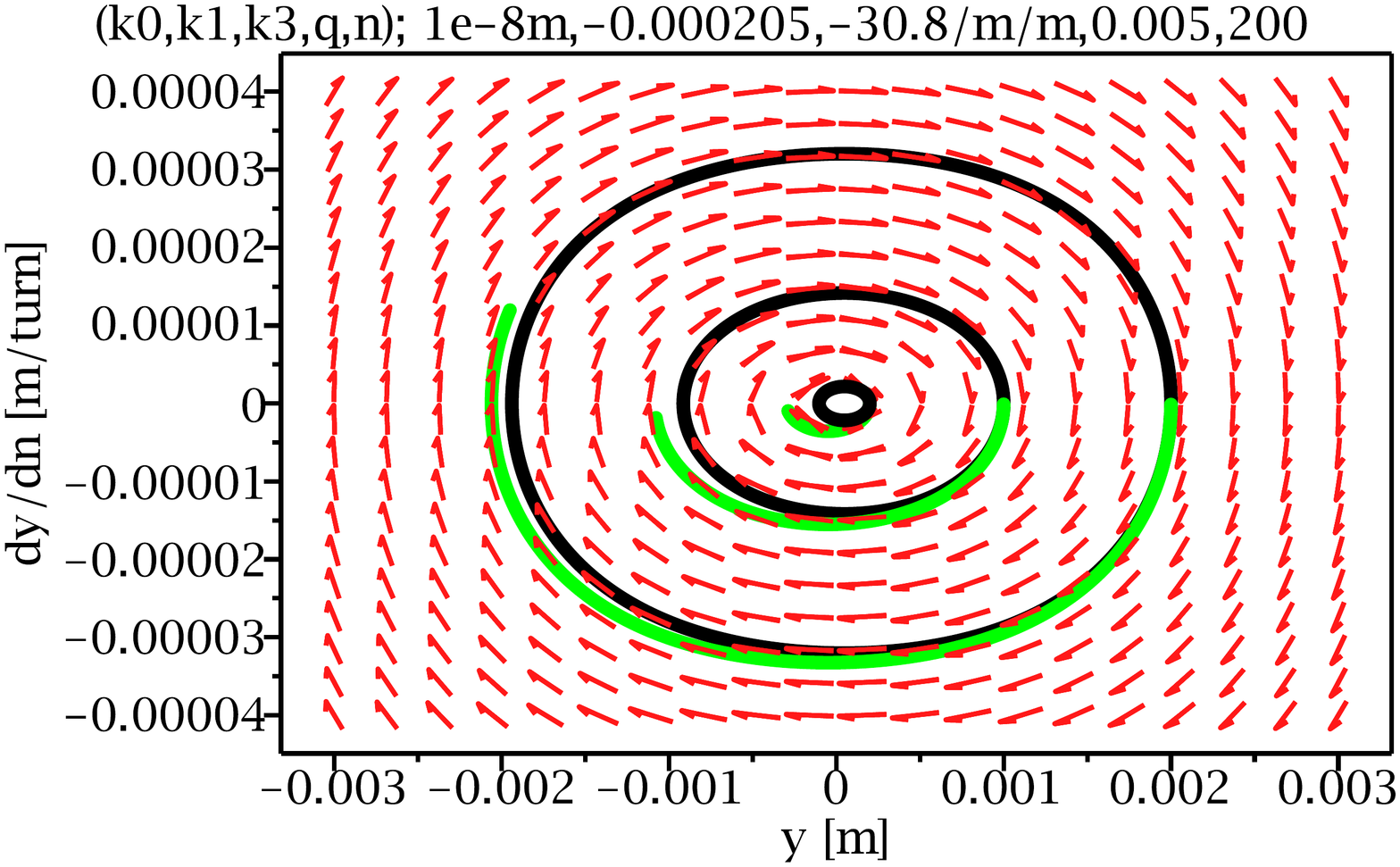}
\caption{\label{fig:PhSpSmallAmpl}From bottom to top, 
small, medium, and large betatron amplitude phase space orbits in relativistic 
self-magnetometer electric storage ring ``bottle'' with only octupole 
focusing. Green orbits are CW, black CCW. Focusing parameters and number 
of turns are shown in the headings. Values of ``radial magnetic field
error coefficient'' $k_0$ and number of turns $n$ are chosen
just large enough to give visible effect in these graphs.
The plots were obtained using MAPLE.}
\end{figure}

With its parameters other than $k_0$
now determined at least semi-quantitatively, Eq.~(\ref{eq:Magnetometer.1}) 
can be investigated numerically. This is done in 
Figure~\ref{fig:PhSpSmallAmpl}, for vertical betatron amplitudes
ranging from 100 microns to 6 cm. These give phase space plots for
CW and CCW orbits. Radial magnetic field error $\langle\Delta B_r\rangle$
(listed above the graphs, and different for the different graphs) 
shift the green orbits relative to the black.

Two parameters influence the magnetometry sensitivity for the EDM measurement
application. One is the r.m.s. accuracy $\sigma_{y,\rm shift}$ of the 
measurement of the orbit shift in switching from CW to CCW beam direction. 
Individual BPM r.m.s. precisions of $10^{-6}$\,m have been achieved 
in modern storage ring 
operations\cite{LiberaBPM}\cite{PrecisionBPM}
\cite{EBPM-XBPMcorrelation}. A design study for a cooling ring for
the ILC collider with similar BPM requirements is described in
reference\cite{CornellILC}. Averaging over BPM's distributed 
around the ring can reduce the error by perhaps a factor of 
10 to produce $\langle\sigma_y\rangle\approx10^{-7}\,$m r.m.s. BPM 
precision.

With simultaneously counter-circulating beams the separation 
uncertainty $\sigma_{y,\rm shift}$, can be considerably smaller,
for example by precise squid measurement of the generated magnetic 
field off to the side\cite{BNLproposal}. Alternatively,
with directional BPM pickups, the beam separation can be measured 
differentially.

The other parameter determining magnetometer sensitivity
is the (dimensionless) ratio 
$(y_{\rm CW}-y_{\rm CCW}/k_0$ giving the centroid orbit reversal shift
divided by $k_0$. Figure~\ref{fig:PhSpSmallAmpl} shows that the ratio 
$(y_{\rm CW}-y_{\rm CCW}/k_0$ ranges from $10^4$ at small amplitude to
roughly $10^3$ at large amplitude. As a result the measured 
beam displacement will depend on the actual vertical beam distribution,
though not very sensitively for beam distributions that are more or
less constant. We assume $(y_{\rm CW}-y_{\rm CCW}/k_0\approx10^4$ which
is probably conservative, considering that the sensitivity can be
increased by adjusting $q$.

The EDM systematic error associated with radial magnetic field
uncertainty can be expressed as a 
number of standard deviations $N_{\tilde d}$, in units of the nominal
EDM value of $10^{-29}$\,e-cm. The result is
\begin{align}
N_{\tilde d}
 &=
\frac{1}{\sigma_y}\,
\frac{y_{\rm CW}-y_{\rm CCW}}{k_0}\,
k_0^{\rm nom}, \notag \\
 &\approx
\frac{1}{10^{-7}\,{\rm m}}\,\times 10^4\times(0.5\times10^{-12}) 
 = 
0.05.
\label{eq:Magnetometer.7}
\end{align}
By this estimate the r.m.s error corresponding to a single 
$\langle B_r\rangle$ compensation would be $2\times10^{-28}\,$e-cm, 
twenty times larger than the nominal EDM value of $10^{-29}$\,e-cm.
Of course the measurement accuracy would be improved by
averaging over multiple runs. 

The fractional error in measuring the effect of $k_0=10^{-8}$\,m 
is the same as the fractional
error in measuring the vertical beam displacement
for $k_0=10^{-8}$\,m, which is 
is $\sigma_y/\Delta y=10^{-7}\,{\rm m}/10^{-4}\,{\rm m}=10^{-3}$. 
So the r.m.s. error in measuring $k_0$ is 
$\sigma_{k0}=10^{-8}\times10^{-3}=10^{-11}$\,m. For the numerical
values given in Table~\ref{tbl:BottleParameters},
$k_0=2.8\times10^4\Delta B_r[{\rm T}]$. From these estimates 
we obtain, for the r.m.s. average magnetic field error,
\begin{equation}
\sigma_{B_r}
 = 
\frac{10^{-11}}{2.8\times10^4}
 =
3\times10^{-16}\,{\rm T}.
\label{eq:Magnetometer.6p}
\end{equation}

\section{Electric Storage Ring Bottle\label{sect:ElectricBottle}}
A quite simple analysis of an idealized relativistic electric storage 
ring bottle has been given. A real storage ring deviates significantly
from this idealization. Perhaps the most significant deviation concerns
the effect of the RF cavity necessarily present for bunched beam operation
of the storage ring. Even if quite weak, the RF cavity constrains the
revolution frequency of every particle to be the same, irrespective of
its betatron amplitudes. No such constraint has been included in the
analytic formulation given so far. Without the RF, revolution frequencies
would depend on betatron amplitudes and fractional momentum 
offset $\delta$, causing unacceptably small spin coherence time (SCT).

There are other deviations from ideal. Real lattices have straight sections.  
In the simulations making up the remainder of this paper, the circumference 
taken up by straight sections is about 15 percent of the total circumference, 
disributed uniformly. This leaves the ring essentially circular, quite faithful
to the analysis so far. The lattice for a real EDM experiment might, more likely, 
be racetrack shaped. Also the need for polarimeters, octupoles, Wien filters, 
beam position monitors, injection regions, and diagnostic equipment of
various sorts, is likely to require a considerably less symmetric ring,
with a smaller fraction of the circumference dedicated to electric bends.

Because the bending elements are discrete, there are end effects in a
real storage ring, which have not been considered so far. As it happens, end 
effects are less serious for orbits than for spin evolution, which is not 
being considered in this paper. In any case, fringe field effects are included 
in our simulations and are found to be unimportant.

The simulation code ETEAPOT, which models all-electric rings,
has been used. Only a limited simulation of the performance of the relativistic 
electric bottle has been attempted. Four starting conditions are listed in 
Table~\ref{tbl:ElectricBottle}, in columns labeled 
tiny, small, medium, and large amplitude radial, vertical, and fractional 
momentum offset $\delta$ for the tracked particles. Note
that, to limit the number of plots displayed, $x$, $y$, and $\delta$ 
amplitudes are scaled more or less proportionally---for example, tiny 
amplitude in, say $x$, and large amplitude in, say $\delta$, is \emph{not}
displayed. Even with this restriction there is an embarrassing proliferation of
figures. Furthermore, the same graphs, with the same initial conditions, are 
repeated later for a magnetic bottle.

\begin{figure*}[h]
\centering
\includegraphics[scale=0.52]{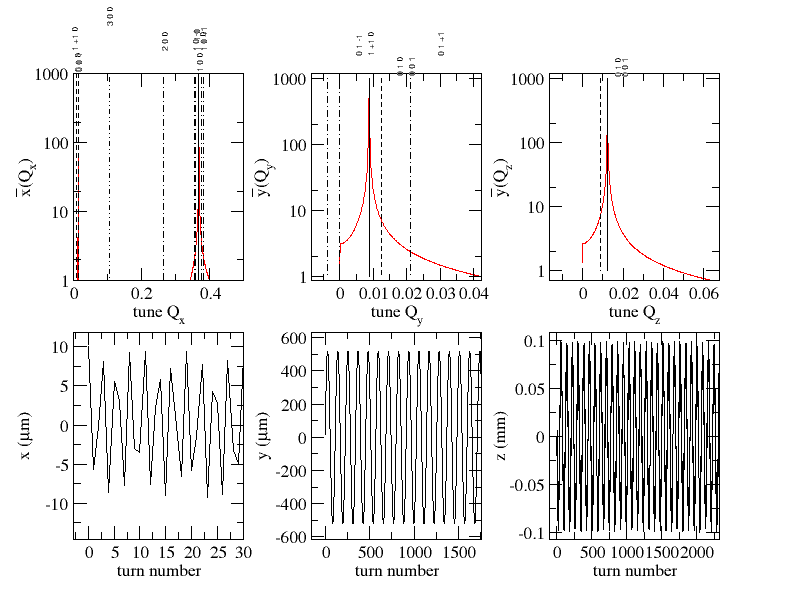}
\includegraphics[scale=0.52]{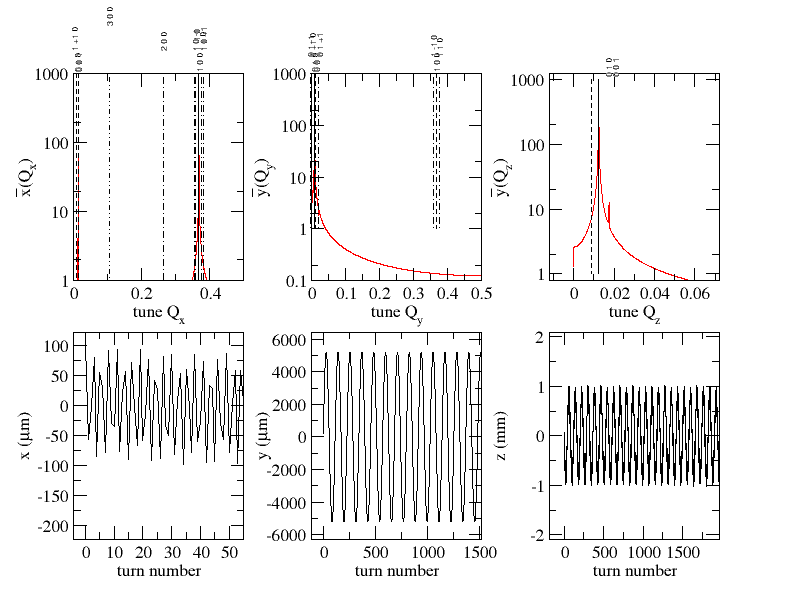}
\caption{\label{fig:elec_smsm_sm.png}
ETEAPOT calculated, {\bf electric bottle}, 
$x$, $y$, and $z$ space and frequency 
domain evolution for ``tiny'' (above) and 
``small'' amplitude (below). In these and all following
spectrum plots, because of aliasing, only fractional
tunes are plotted. In every case the integer tune values
are 1 for $Q_x$ and 0 for $Q_y$ and $Q_z$.
}
\end{figure*}

\begin{figure*}[h]
\centering
\includegraphics[scale=0.54]{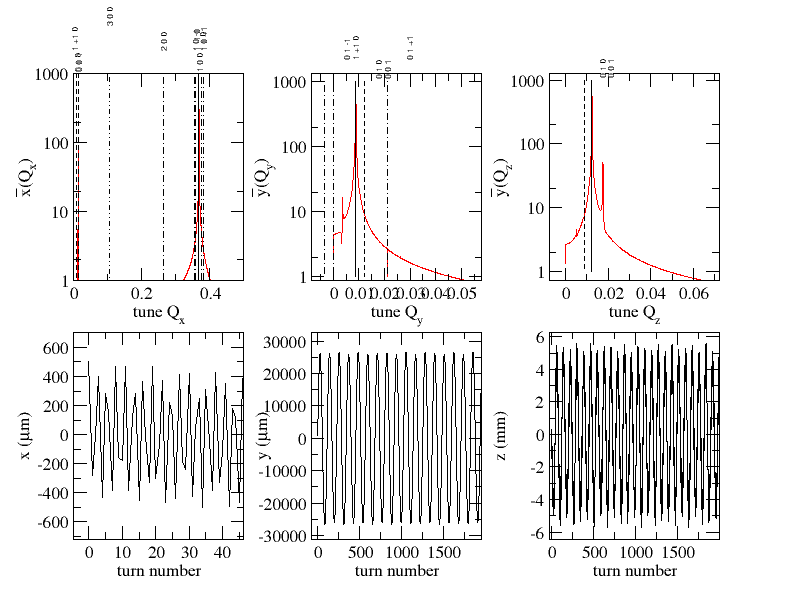}
\includegraphics[scale=0.54]{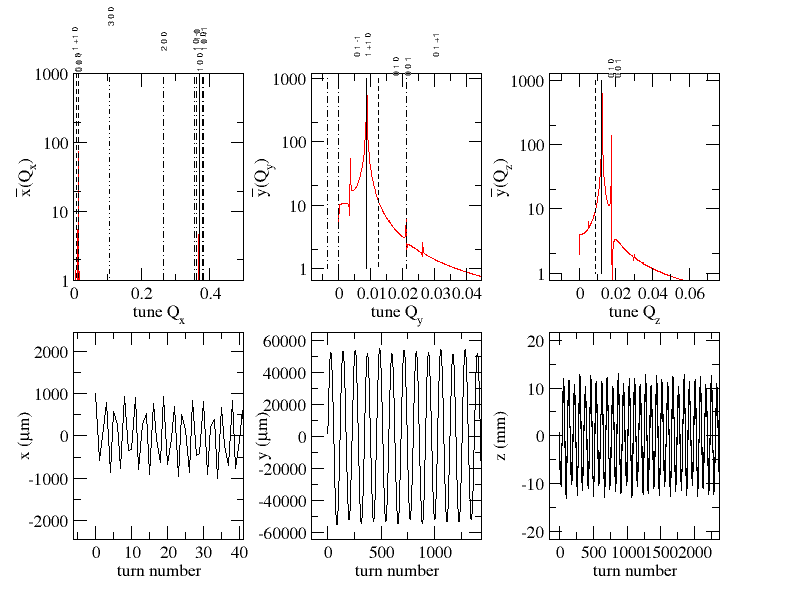}
\caption{\label{fig:elec_big.png}ETEAPOT calculated, 
{\bf electric bottle}, $x$, $y$, and $z$ space and frequency 
domain evolution for ``medium'' and ``large'' amplitude. }
\end{figure*}

The revolution frequency for this ring is $f_0=0.6c/281.4=0.6897\,$MHz.
Only one (quite low) RF voltage drop $V_{rf}=50\,$keV has been investigated.
The RF harmonic number is $h=10$, making $f_{rf}=6.897$MHz. (Much higher
harmonic number and lower voltage might be more practical.) These parameters 
fix the lengths of beam bunches (which can be estimated from the $z$ vs turn
number plots).
\begin{table}[h] 
\begin{tabular}{|c|c|c|c|c|c|}  \hline
amplitude &   unit  &    tiny    &   small  &   medium  &   large   \\ \hline
$x_0$     &    m    &  1.0e-5    &  1.0e-4  &   0.5e-3  &  1.0e-3   \\
$x'_0$    &         &     0      &    0     &     0     &    0      \\
$y_0$     &    m    &  2.0e-5    &  2.0e-4  &   1.0e-3  &  2.0e-3   \\
$y'_0$    &         &  1.0e-7    &  1.0e-6  &   0.5e-5  &  1.0e-5   \\
$\delta$  &         &  1.0e-8    &  1.0e-7  &   0.5e-6  &  1.0e-6   \\ \hline
$\hat x$  &   cm    &  0.001     &   0.01   &   0.05    &    0.1    \\
$\hat y$  &   cm    &  0.05      &   0.50   &   2.50    &    5.0    \\ \hline
$Q_x$     &         &  1.369     &   1.371  &   1.369   &   1.371   \\
$Q_y$     &         &  0.0091    &   0.0091 &   0.0088  &   0.0090   \\
$Q_s$     &         &  0.0125    &   0.0125 &   0.0118  &   0.0125   \\ \hline
\end{tabular}
\medskip
\caption{\label{tbl:ElectricBottle}Parameter dependencies, ranging from small
to large amplitude particles, tracked in {\bf electric bottle}.}
\end{table}
Especially because there are no quadrupoles, nor other linear focusing,
the most important question to be established is whether oscillations in
all three degrees of freedom are stable. Sinusoidal motion for brief time 
intervals, i.e. small numbers of turns, provides necessary, though not sufficient, 
evidence of long time stability. All of the time 
series graphs, for all amplitudes of all three degrees of freedom, satisfy 
this test. These are the lower plots in Figures~\ref{fig:elec_smsm_sm.png} 
and \ref{fig:elec_big.png}.
Note that, to make the sinusoidal oscillations visible, some graphs have 
been zoomed to show fairly small numbers of oscillations. In all cases the
maximum excursions neither shrink nor grow over long times, indicating
stability.

A more stringent test of long term stability is exhibited in the upper
figures, which provide spectral analyses of the full data sets for the
lower figures. The narrow frequency spectra demonstate stability in all
cases. Again some of the spectra have been zoomed to permit the line 
centroids to
be determined accurately. The solid vertical lines are centered on these
lines, and it is these tunes that are listed as $Q_x$, $Q_y$, and $Q_s$
in Table~\ref{tbl:ElectricBottle}. These are fractional tunes. 
``Aliasing'' suppresses integer tunes. As it happens this affects only
$Q_x$; its integer tune of 1 has been included in the table.

The broken vertical lines in the tune spectrum plots locate likely
nonlinear resonance tunes. In all the plots, only a couple
of matches are observed, and they have small amplitudes. This should
excuse the absence of explanation of the cryptic (but actually simple) 
index labels printed above the graphs. Complicating the explanation would 
be the fact that the alignment of the indices with the broken lines
(which are located properly) has been messed up in those plots that have 
been zoomed.

At even larger amplitudes than are exhibited in these
graphs nonlinear resonance proliferate at their expected frequencies
and, ultimately, the motion becomes chaotic, limiting the dynamic
aperture.

As expected the longitudinal tune $Q_s$ is independent of all amplitudes. 
Dependence of $Q_x$ and $Q_y$ on amplitude is exhibited by plotting values 
from the table in Figure~\ref{fig:ElecMagn-x}. The expected 
constancy of $Q_x$ and the strong
dependence of $Q_y$ on vertical amplitude is observed---$Q_y$ is
more or less proportional to amplitude. The only important effect
of octupole-only focusing is to alter the $Q_y$ dependence.

The graphs in Figure~\ref{fig:ElecMagn-x} also show the corresponding 
tune dependencies for the
purely magnetic storage ring investigated in the next section.
With conventional quadrupole focusing both $Q_x$ and $Q_y$ are
independent of betatron amplitudes. 

\section{Magnetic Storage Ring Bottle\label{sect:MagneticBottle}}
In this section a magnet lattice, differing from the previously
discussed electric lattice primarily by the replacement of electric
bends by uniform field magnetic bends, is investigated. Like the electric
lattice, this lattice would, without vertical focusing, be vertically unstable. 
In this case, vertical stability is imposed by discrete vertically 
focusing quadrupoles, situated between bends, and just barely strong 
enough to provide vertical stability. Tracking is performed using
UAL/TEAPOT\cite{TEAPOT}\cite{UAL}.

\begin{figure*}[h]
\centering
\includegraphics[scale=0.55]{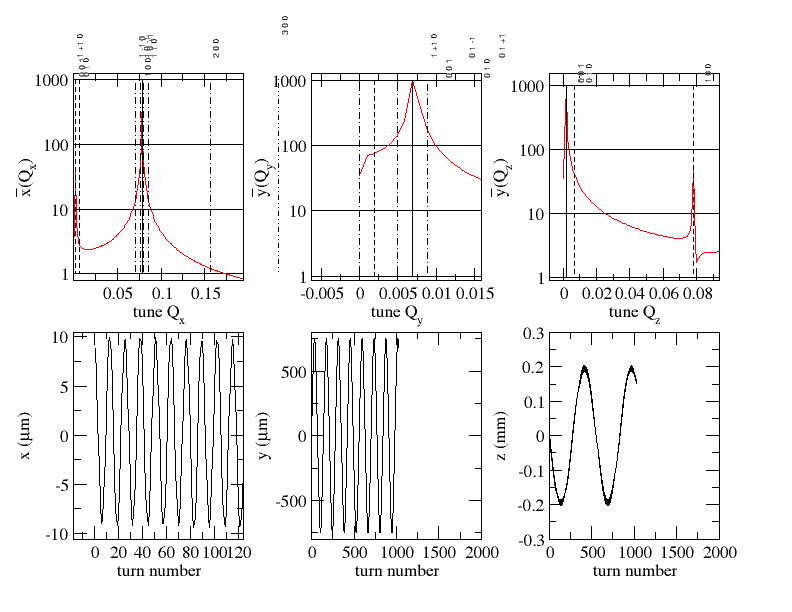}
\includegraphics[scale=0.55]{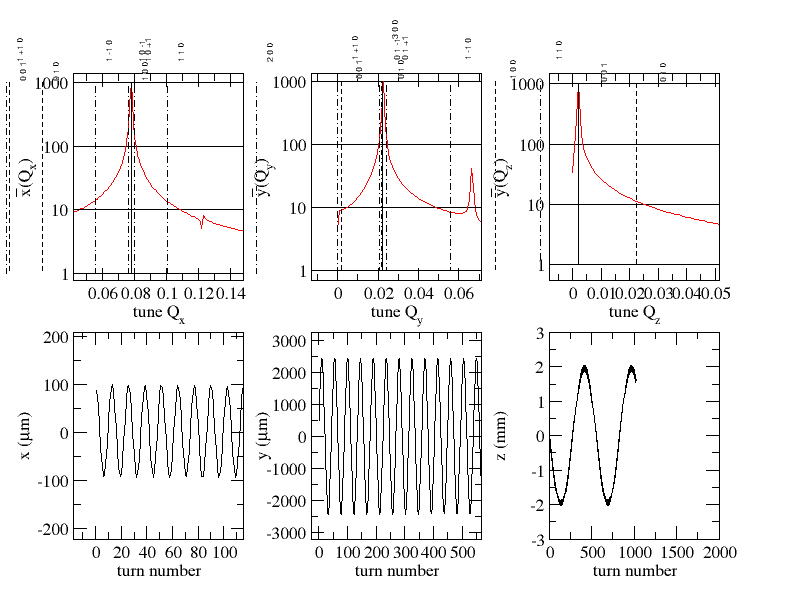}
\caption{\label{fig:mag_smsm.png}TEAPOT calculated, 
{\bf magnetic bottle}, $x$, $y$, and $z$ space and frequency 
domain evolution for ``tiny'' (above) and 
``small'' amplitude (below). }
\end{figure*}

\begin{figure*}[h]
\centering
\includegraphics[scale=0.55]{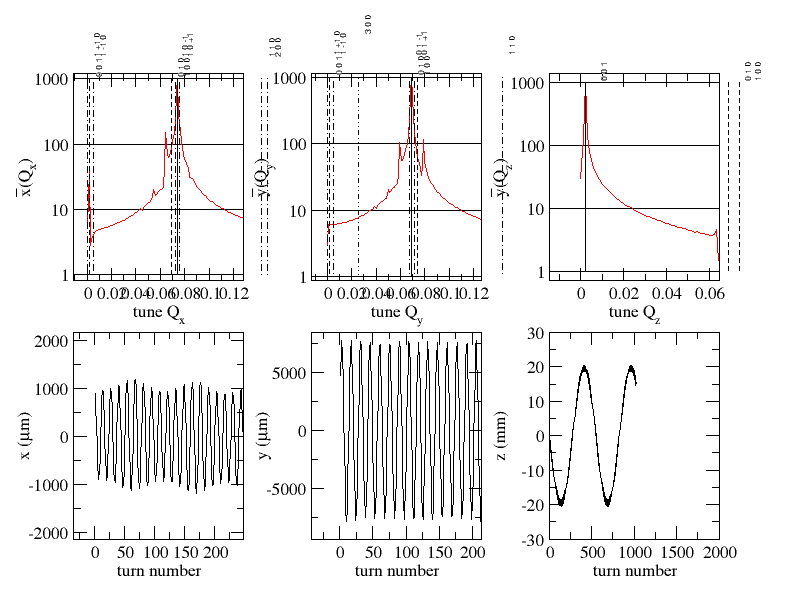}
\includegraphics[scale=0.55]{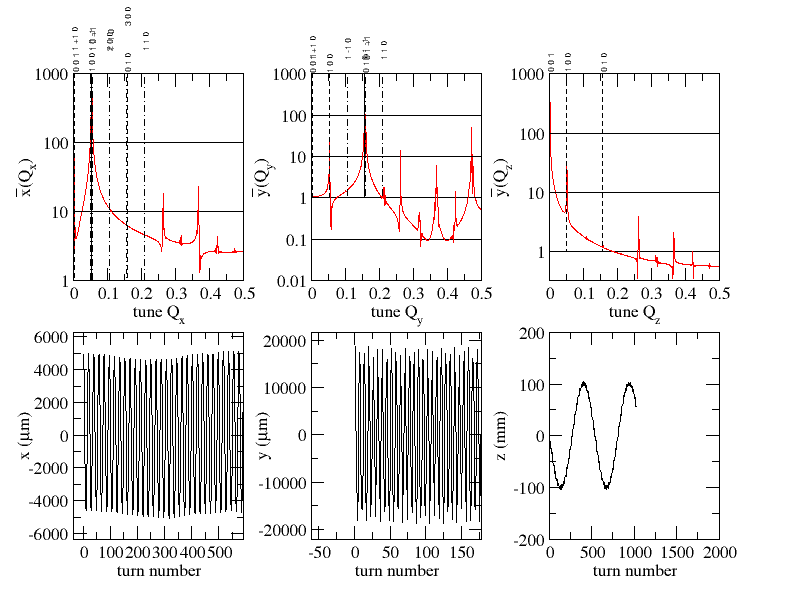}
\caption{\label{fig:mag_med.png}TEAPOT calculated, 
{\bf magnetic bottle}, $x$, $y$, and $z$ space and 
frequency domain evolution for ``medium'' (above) and 
``large'' amplitude (below).  }
\end{figure*}

Particles with more or less the same tiny, small, and medium
initial amplitudes as in the previous section are tracked, 
and the results are shown in the same
sequence of plots, in Figures~\ref{fig:mag_smsm.png} and 
\ref{fig:mag_med.png}. However the ``large'' amplitude initial 
conditions that are stable in the electric ring are unstable in
the magnetic ring. So the ``large'' amplitude initial conditions
were reduced by a factor of two for the magnetic ring. Even so, 
insipient nonlinear behavior can be noted, by the proliferation of
extra lines in the $Q_y$ tune spectrum for the large amplitude
case. The small peak identified by the broken 
line labeled ``1 1 0'' in the $Q_y$ tune plot
has a frequency indicating that it is due
to nonlinear $(x,y)$ resonance. The stronger peripheral lines
are probably also due to nonlinear resonance, but their frequencies
are not identified to confirm this.

\begin{figure*}[h]
\centering
\includegraphics[scale=0.7]{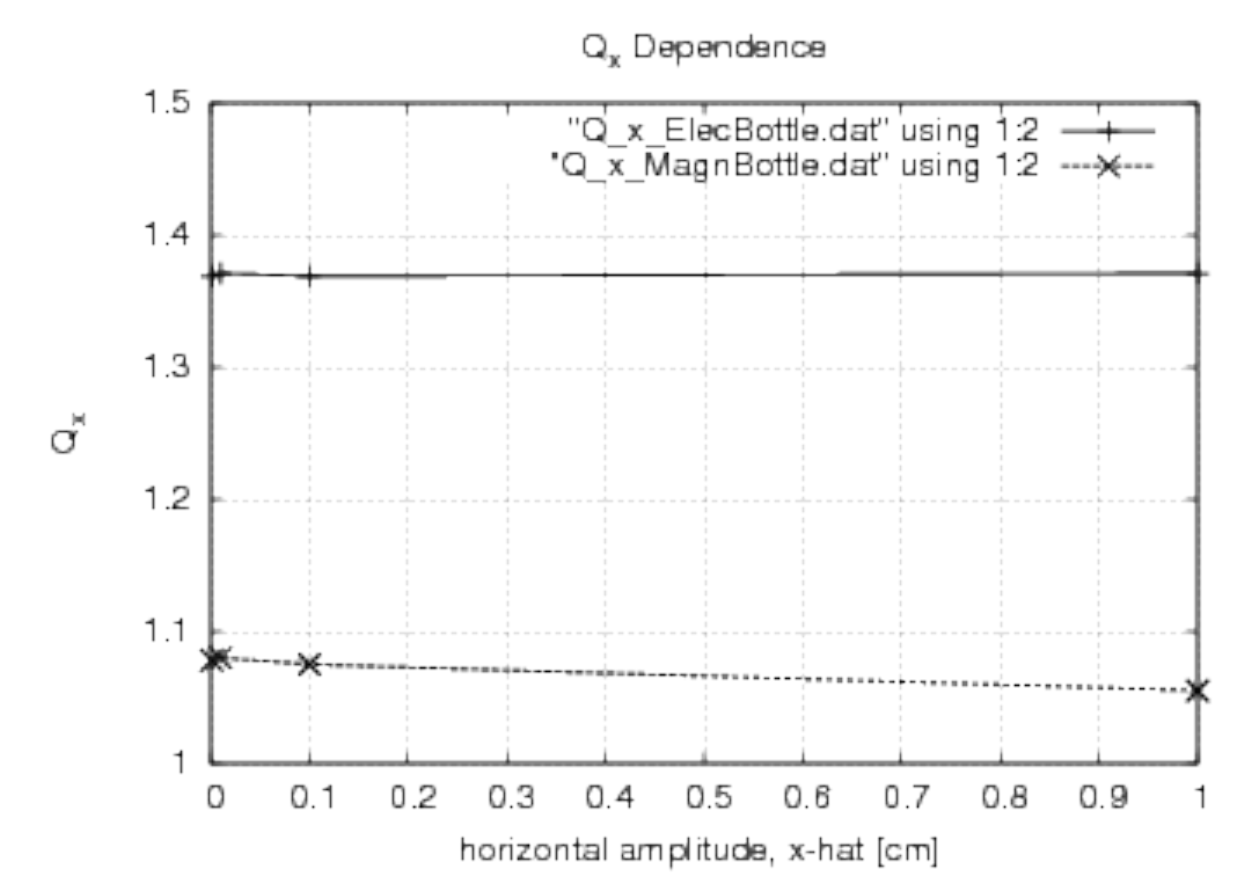}
\includegraphics[scale=0.7]{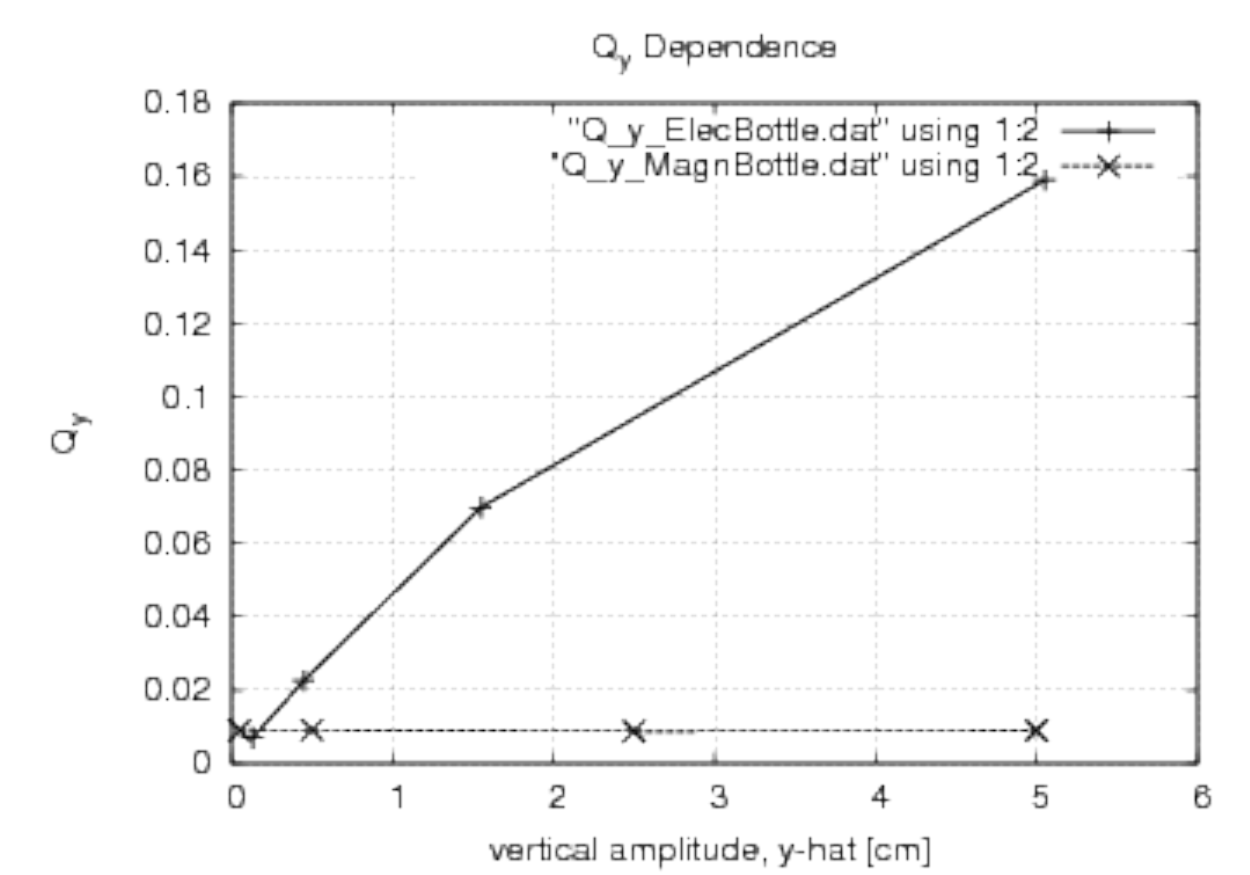}
\caption{\label{fig:ElecMagn-x}Horizontal and vertical 
tunes in electric and magnetic relativistic storage ring bottles. 
With, at most, very weak explicit quadrupole focusing elements, the 
horizontal electric ring tune $Q_x$ is not far from value 1.28,
the pure circular ring theoretical value. For the magnetic ring 
the horizontal tune $Q_x$ is not far from 1.0, the no
quadrupole theoretical value. The only strikingly novel feature 
is the near-linear dependence of $Q_y$ on vertical amplitude 
in the electric case. Of course this is due to the all-octupole 
focusing.
}
\end{figure*}

The magnetic lattice graphs are subject
to pretty much the same discussion as for the electric lattice.
Amplitudes and tunes extracted from the graphs
are entered in Table~\ref{tbl:MagnticBottle}. 
The dependence of tunes on amplitude are plotted in 
Figure~\ref{fig:ElecMagn-x}, where
they can be compared to the corresponding tunes in the electric ring.
Since this is a conventional storage ring both $Q_x$ and $Q_y$ are
constant, independent of amplitude.
\begin{table}[h] 
\begin{tabular}{|c|c|c|c|c|c|}  \hline
amplitude &   unit  &    tiny    &   small  &   medium  &   large   \\ \hline
$x_0$     &    m    &  1.0e-5    &  1.0e-4  &   1.0e-3  &   0.5e-2  \\
$x'_0$    &         &     0      &    0     &     0     &      0    \\
$y_0$     &    m    &  2.0e-5    &  2.0e-4  &   2.0e-3  &  1.0e-2   \\
$y'_0$    &         &  1.0e-7    &  1.0e-6  &   1.0e-5  &  0.5e-2   \\
$\delta$  &         &  1.0e-8    &  1.0e-7  &   1.0e-6  &  0.5e-5   \\ \hline
$\hat x$  &   cm    &  0.001     &  0.010   &   0.10    &  1.0      \\
$\hat y$  &   cm    &  0.130     &  0.450   &   1.55    &  5.05     \\ \hline
$Q_x$     &         &  1.078     &  1.080   &   1.075   &  1.055    \\
$Q_y$     &         &  0.0069    &  0.0226  &  0.0698   &  0.159    \\
$Q_s$     &         &  0.0021    &  0.0022  &  0.0020   &  0.0021   \\ \hline
\end{tabular}
\caption{\label{tbl:MagnticBottle}Parameter dependencies, ranging from small
to large amplitude particles, tracked in {\bf magnetic bottle}. $Q_s$
appears to depend on amplitude, but not outside error bars which are not shown. }
\end{table}

\section{Recapitulation and Conclusions\label{sect:Recapitualtion}}
\subsection{Motivation}
The novel relativistic bottles proposed here are more than a 
curiosity. Especially the electric bottle is appropriate for 
measuring the electric dipole moments, especially of the proton, 
but also the electron and other fundamental particles. Because 
of the proton's particular anomalous magnetic moment, a polarized 
proton beam of ``magic'' 234\,MeV kinetic energy, can be ``frozen'', 
meaning that the polarization vector rotates at the same rate, and 
around the same axis, as the momentum. Out-of-plane tipping
of the beam polarization provides the signal basic to the
storage ring measurement of the proton EDM. 
(See reference\cite{BNLproposal}.)

EDM measurement relies on measuring the precession caused by
the electric field acting on the particle EDM. But the proton
MDM is vastly greater than the EDM (whose very existence implies
breakdown of both parity and time reversal invariance) for which
only upper limits are known. By design $B_r=0$ but, inevitably,
there will be $\Delta B_r$ error fields.
The only way to minimize MDM-induced
precession is to cancel the magnetic field, primarily by
magnetic shielding, secondarily, by cancelling residual magnetic
fields by compensation coils. 

The achievable EDM accuracy depends on the precision with which
the magnetic field can be zeroed.  Fortunately it is only the
magnetic field averaged over the design orbit that has to 
be cancelled. Self-magnetometry provides the ideal magnetic
measurement for achieving this end. This is because a shift of 
the beam orbit caused by the unknown magnetic field error, 
is strictly proportional to the spin precession that needs 
to be cancelled. 

The all-electic case, appropriate only for proton or electron
EDM measurement, has been emphasized in this paper. This is the
only case where counter-circulating beams can co-exist. However
the high-sensitivity magnetometry will be applicable also
for rings containing both magnetic and electric bending.

\subsection{Applicability of Electric Bottle}
Conventional storage rings rely on quadrupole focusing. This paper 
has shown how storage rings with only octupole focusing
elements will perform much like conventional rings.
It is ``sefl-magnetometry'' that motivates replacing quadrupole 
focusing by octupole focusing. The electric storage ring bottle 
with only octupole focusing is ideal for this task. Because the 
orbit shift to be detected is vertical, it is important for the 
vertical focusing to be \emph{weak}---a small vertical force gives 
a large vertical orbit shift. Relative to this focusing, the
geometric horizontal focusing associated with the uniform
bending field is strong.
As a result, horizontal motion is much the same as in a conventional
ring. It is only the vertical focusing that is weak. This contrast
is most clearly visible in Figure~\ref{fig:ElecMagn-x}.  
In particular, unlike in a conventional
ring, the vertical tune $Q_y$ is roughly proportional to the vertical
betatron amplitude.

\subsection{Projected Operational Performance}
The performance of an all-electric relativistic storage ring with only 
octupole focusing has been investigated in 
Section~\ref{sect:ElectricBottle} from the point of view of regular
storage ring operation. There is satisfactory stability in all three 
phase space coordinates in spite of the fact that vertical focusing is
provided by octupoles rather than quadrupoles. Furthermore, the
geometric focusing in the bend elements provides ample horizontal
focusing. In short, no quadrupoles (except, perhaps, trim quadrupoles) 
need to be present in the lattice. 

In  Section~\ref{sect:MagneticBottle} a quite similar magnetic ring is 
analysed and found to perform similarly.
The Section~\ref{sect:Appendix} appendix gives a formalism, 
copied from well known plasma physics theory, for describing a
storage ring as a magnetic bottle. The purpose for this appendix 
is to emphasize that charged particles storage in a ``bottle'' is 
not new, and that the nonrelativistic theory can easily be applied
to relativistic storage rings. 

\subsection{Proton EDM Measurement Options\label{sect:EDMOptions}}
This section is something of a epilogue, contemplating the 
experimental implications for the proton EDM measurement 
that motivated the paper. The main EDM measurement issues
were introduced in Section~\ref{sect:Magnetometer}. The variables
are: electron or proton; resonant polarimetry or scattering polarimetry;
simultantaneously circulating beams, or one-at-a-time beams? 
Koop rolling spin wheel or not?; what about the Wien filter 
directional sensitivity? etc. Far too many for this brief concluding 
section. Fortunately, other than the essential difference 
between electron and proton rings, the same lattice designs will 
be applicable for all choices among the variables. 
Here I will fix on the proton case, and the two configurations 
I currently consider most promising.

Though not mentioned so far, and not proven so far, the spin 
coherence time performance of the octupole-only ring is likely to 
be superior to any of the rings studied to date. This is because 
the lattice is closest to the pure $m=0$ cylindrical lattice for 
which, theoretically, the SCT is infinite. The only blemish 
needing further study concerns the spin decoherence 
caused by the (very weak) octupole field.

More important, because the $\langle B_r\rangle$ spurious precession 
is the most serious source of systematic EDM error, the octupole-only 
bottle storage ring, invented for the purpose, and described in this
paper, seems to me to be obligatory. Also
for sufficiently high precision, digital frequency domain
resonant polarimetry is obligatory. But resonant polarimetry 
requires the Koop spin wheel rolling polarization, and \emph{rolling 
polarization requires a local Wien filter which, because of its 
directionality, may rule out simultaneously counter-circulating beams.} 
This seems to reduce the options down to single beams, alternating between
CW and CCW runs, with $\langle B_r\rangle$ compensation between
runs. Aspects of this design have been considered in 
Section~\ref{sect:Magnetometer}. This EDM experimental route has 
the potential for measuring a proton EDM value as small, 
let us say\cite{FreqDomainEDM}, as $10^{-29}\,$e-cm, probably
limited by the systematic error caused by spurious MDM-induced
precession in unknown residual radial magnetic field.

The only known way to further reduce this systematic error is to have 
simultaneously counter-circulating beams---an option tentatively
ruled out in the italicized sentence in the previous paragraph. 

The problem is that controlling the rolling polarization of even 
a single beam requires two Wien filters. 
The Wien filter with cancelling horizontal deflections 
is needed to keep the spin wheel of the polarized
beam properly aligned without affecting the
orbit at all because the electric and magnetic deflections
cancel. But, acting on the counter-circulating beam,
the electric and magnetic deflections add. Fortunately 
even summed, this kick is a mere ``tickle''. This Wien
filter is only cancelling the spin effects of unknown 
tiny radial deflections that, nominally, average to zero.
Furthermore, any beam growth or change in 
particle distribution induced in a counter-circulating 
beam by these kicks is horizontal. 
As such it has little or no tendency to introduce up-down 
asymmetry that could influence the EDM measurement. 
In spite of its directionality, this Wien filter application
is therefore probably harmless.

The Wien filter with cancelling vertical electric
and magnetic deflections cannot be taken so lightly. The
spin kicks from this Wien filter need to be at least strong
enough to drive the Koop spin wheel
polarization roll, but without influencing the polarized beam 
orbits. Furthermore, \emph{precise reversal} of the polarization roll is essential
for obtaining the EDM measurement. This spin wheel reversal,
if applied by a Wien filter, applies 
an uncompensated local vertical 
kick to the counter-circulating beam orbit, possibly
producing effects mimicking the effect of particle EDM. 
This is not good. 

I can think of only one way to fix this problem.
It is to use the global $\langle B_r^{\rm comp}\rangle$ 
circuit, rather than a local Wien filter, to impose 
the rolling polarization. 

It has been explained in 
Section~\ref{sect:Magnetometer} how,
working with the $\langle E_y^{\rm comp}\rangle$ compensation, 
the overall radial magnetic field average can be nulled
with both beams centered vertically.
From this perfectly balanced condition, a tiny shift 
of $\langle B_r^{\rm comp}\rangle$, adiabatically applied, 
will introduce the required polarization roll.
However a tiny vertical beam separation beween the
counter-circulating beams will also
occur; it is an inevitable consequence of the radial
magnetic field driving the roll. This separates the
beam centroids everywhere in the ring. But the shift is
miniscule and will be treated by correction. 

(Vertical separation introduces a small systematic vertical 
``force'' $\Delta f_y$ on one beam due to the other but, 
because the separation is small compared to the beam height, 
this will not alter the orbits significantly. \emph{There will, however,
need to be a systematic correction for the MDM-induced spin 
precession due to the magnetic field of the other beam.})

Otherwise there will
be no significant perturbation of either counter-circulating
beam distribution. This is because, unlike a local
Wien filter, the compensation circuit is global,
and varies only adiabatically. For example, 
half way through each run $\langle B_r^{\rm comp}\rangle$
will be slowly reversed as part of the EDM measurement
sequence. I conjecture that any spurious systematic EDM 
signal caused by this sequence of operations can be
corrected analytically with satisfactory accuracy.

Another issue to be faced is the ``incompatibility'' of  
simultaneously counter-circulating beams with resonant 
polarimetry (which is itself delicate and, as yet, unproven).
The resonant polarimeter is a highly tuned
device whose response is only made detectible after millions of 
coherent bunch passages. The possiblility of separation bumps 
enabling the beams to pass through separate polarimeters has 
been contemplated. But this brings in complications too horrible 
to mention. So I take it as given that both beams have to pass 
through the same resonators and, in fact, the same everything!

Two beams passing through a single resonator certainly represents a
complication. Another complication comes from the fact that
each beam has many bunches, presumed to be equally spaced; 
this is not new, but it is essential to the discussion. To simplify
the discussion the rolling spin operation will be turned
off for this discussion, returning to truly frozen spin operation,
parallel, or anti-parallel to the beam orbits.

Ideally one would wish to monitor and phase lock the beam
polarizations of the two beams independently, but this is 
probably impossible. The beams  
are injected with independent errors, e.g. different energies,
they pass through all the same elements, and there is no 
significant damping mechanism. The resonator itself cannot 
distinguish between the beams. Satisfying superposition, 
the resonator linearly superimposes the signals from all passing 
bunches, irrespective of their directions of travel. So resonant 
polarimeters in the ring simply register the coherent sum of 
the two beam polarizations. 

This coherent summing of the two beam polarizations may or may
not be tolerable. But, even if not tolerable, there is a workaround. 
Only one or the other of the CW and CCW beams needs to be 
polarized for the EDM measurement. 
An unpolarized beam applies no magnetization signal
at all to the polarimeter. So, running with one 
highly-polarized beam and one unpolarized beam, the polarimeter 
responds only to the polarized beam, as if there is a 
single beam. The unpolarized beam is ``inert''.

The unpolarized beam provides no EDM information. Its
role is to facilitate the self-magnetometry. But it
may also be possible for the inert beam to play another role.
If the two beam currents are exactly equal (which is not
otherwise essential for self-magnetometry) the net current 
through the magnetometer vanishes. It may be possible to 
take advantage of this to cancel the direct excitation of 
the resonator by the circulating beam currents. This would
greatly relax one of the serious uncertainties concerning
resonant polarimetry. This could, for example, permit
the roll polarization frequency to be reduced from, say, 
100\,Hz to 1\,Hz, while still permitting the polarization 
frequency line to be resolved from the nearby revolution 
harmonic.

It is not clear whether this will work; direct 
resonator excitation is more feared than understood
at this time. 
Like the magnetization response, the direct response
is also the coherent sum of the responses to the
separate beams. The opposite beam directions and
the phase shifts through the resonator make it
possible for the resonator to be excited even when
the net beam current vanishes. By symmetry, arranging the 
resonator center to coincide with bunch crossing points 
will probably be either optimal or anti-optimal; i.e. the 
superposition is either perfectly constructive 
or perfectly destructive. If anti-optimal, then placing 
the resonator midway between bunch crossings would be 
optimal. One only has to design the ring lattice
and the RF phasing appropriately.

From this point of view counter-circulating
beams may even be helpful for resonant polarimetry. 
Certainly, this possibility requires further study.

Incidentally, it is already known\cite{BNLproposal}, 
that beam-beam interaction of the counter-circulating 
beams has negligible detrimental effect on storage 
ring EDM measurement.

All this implies that simultaneously counter-circulating
beams may be practical after all. This route has the potential 
for reducing the EDM systematic error by a substantial
factor---perhaps as much as a factor of ten. This depends 
on the relative seriousness of unknown radial magnetic 
fields and the extent to which problems associated with
having two beams can be mastered.

\clearpage

\begin{center}
      {\bf APPENDIX}
\end{center}

\section{Relativistic Magnetic Bottle\label{sect:Appendix}}
This appendix generalizes to relativistic magnetic traps
formulas from, for example, a book by Spitzer\cite{Spitzer}. 
Results concerning such a relativistic 
magnetic trap are corroborated in the numerical simulations in 
Figures~\ref{fig:mag_smsm.png}, \ref{fig:mag_med.png}, and
\ref{fig:ElecMagn-x}.

Some of the formulas in this appendix 
have been used to include the effects of a very weak 
magnetic field in an otherwise purely electrostatic field 
(not counting RF), which is the primary system investigated in 
the present paper. Behavior of the relativistic all-electric 
trap emphasized in the present paper is qualitatively very similar
to behavior of a magnetic trap.

The dominant field in a magnetic bottle is ${\bf B}=\nabla\times{\bf A}$, 
where the vector potential ${\bf A}$ is given by
\begin{equation}
\Delta A_x=-\frac{1}{2}\,z B,\quad
\Delta A_y=0,\quad
\Delta A_z=\frac{1}{2}\,x B.
\label{eq:MagBottle.1}
\end{equation}
In the present context ``longitudinal'' means vertical or $y$,
``perpendicular'' means horizontal or $(x,z)$.
The relativistic Lagrangian for this motion is given by
\begin{equation}
L = -\frac{mc^2}{\gamma(v)} - e(\phi-{\bf v}\cdot{\bf A}).
\label{eq:MagBottle.2}
\end{equation}
The conjugate momentum vector ${\bf P}$, upper-case to
distinguish it from mechanical momentum (lower-case) ${\bf p}$,
is defined by
\begin{equation}
{\bf P}
 =
\frac{\partial L}{\partial{\bf \dot v}}
 =
{\bf p} + e{\bf A}.
\label{eq:MagBottle.3}
\end{equation}
The Hamiltonian is 
\begin{equation}
H
 = 
{\bf P}\cdot{\bf v} - L
 =
\gamma mc^2 + e\phi
 =
\sqrt{({\bf P}-e{\bf A})^2 +m^2c^2} + e\phi.
\label{eq:MagBottle.4}
\end{equation}
In the approximately uniform magnetic field of a magnetic bottle
with average value $B$ at the orbit position, with value $B_0$ at
the center, the particle
gyrates around a field line with orbit radius  
$r=p_{\perp}/(eB)$. The closed line integral $I_g$ over one turn,
\begin{align}
I_g
 &= 
\frac{1}{4\pi}\,\oint {\bf P}_{\perp}\cdot{\bf dl}_{\perp}
 =
\frac{1}{4\pi}\,\oint\Big({\bf p}_{\perp}+e{\bf A}\Big)\cdot{\bf dl}_{\perp}
\notag\\
 &=
\frac{1}{2}\,p_{\perp}r - \frac{1}{4}eB_0r^2
 =
\frac{1}{4}\,\frac{p^2_{\perp}}{eB}.
\label{eq:MagBottle.5}
\end{align}
is an adiabatic invariant of the gyration, conserved as the
center of the particle's circular orbit moves along its field line.
The single current orbit loop also
has a magnetic moment equal to current times area;
\begin{equation}
\mu_B
 =
e\,\frac{v_{\perp}}{2\pi r}\,\pi r^2
 =
\frac{1}{2}\,ev_{\perp}r
 =
\frac{1}{2}\,e\frac{p_{\perp}}{m\gamma}\,\frac{p_{\perp}}{eB}
 =
\frac{1}{2}\,\frac{p_{\perp}^2}{m B \gamma},
\label{eq:MagBottle.6}
\end{equation}
which differs from $I_g$ only by a constant factor (which does
however include a factor $\gamma$). 
\begin{equation}
\mu_B
 =
e\,\frac{v_{\perp}}{2\pi r}\,\pi r^2
 =
\frac{2}{\gamma}\,I_g.
\label{eq:MagBottle.6p}
\end{equation}
The product
$\mu_BB=I_gB\,2/\gamma$ represents the energy the particle 
by virtue of its magnetic moment in the magnetic field. 

Motion along the field line can be analysed using the constancy 
of $I_g$. Neglecting the longitudinal velocity, 
the kinetic energy of motion in the
perpendicular plane is given by $E_{\perp} = m\gamma c^2$.
But, because of its initial vertical velocity, a particle 
will drift vertically. Since the magnetic field is non-uniform
this will lead it into a region where $B$ is different.
Superficially this seems to contradict the constancy of
$I_g$, since the speed of a particle cannot change in a pure 
magnetic field. It \emph{has to be} that energy is transferred to 
or from motion in the longitudinal direction. We can analyse 
the longitudinal motion on the basis of energy conservation.

When the magnitude of $B$ falls with $I_G$ constant, the 
transverse kinetic energy has to increase. This energy necessarily 
comes from the longitudinal kinetic energy, which falls. Eventually 
a ``turning point'' is reached where the longitudinal energy vanishes.
and the direction of motion of the guiding center along the field
line reverses. 

Even though the total velocity is relativistic, for our
storage ring application, we can assume the vertical
velocity $u_{\parallel}$ can be treated non-relativistically,
provided the particle rest mass is replaced by its inertial mass,
larger by the factor $\gamma$. For purposes of studying its motion
parallel to the magnetic field, 
the total longitudinal energy of a particle is then given by 
the non-relativistic formula
\begin{equation}
h
 = 
\frac{\gamma}{2}\,I_gB(y) + \frac{1}{2}\,m\gamma u^2_{\parallel}.
\label{eq:MagTrap.12}
\end{equation}
Here $h$ is the numerical value of the total energy (kinetic
plus magnetic) of a particular particle being tracked. 
(The same symbol $h$, with a similar meaning is employed
in Section~\ref{sect:FastSlow} to express the (non-relativistic)
vertical energy of a proton in an electric bottle.)
We have specialized from a general case in which the
magnetic field depends only on one coordinate $y$, rather
than a general position ${\bf R}$.
Since the first term depends only on position $y$
it can be interpreted as potential energy. It is larger
at either end of the trap than in the middle. Since both
$E$ and $I_g$ are conserved, this equation can be solved
for the dependence of longitudinal velocity on position $y$;
\begin{equation}
u_{\parallel}(y)
 = 
\pm\sqrt{\frac{2}{m}\,\Big(\frac{h}{\gamma}-\frac{1}{2}\,I_gB(y)\Big)}.
\label{eq:MagTrap.13}
\end{equation}
In a uniform field $u_{\parallel}$ would be constant, but
in a spatially variable field $u_{\parallel}$ 
varies slowly. As the
particle drifts toward the end of the trap, the $B$ field 
becomes stronger and $u_{\parallel}$ becomes less. 
At some value $y_{\rm tp}$ the right hand side of 
Eq.~(\ref{eq:MagTrap.13}) vanishes.
This is therefore a ``turning point'' of the motion,
and the guiding center is turned back to drift toward
the center, and then the other end. Perpetual longitudinal
oscillation follows. But the motion may be far from
simple harmonic, depending as it does on the detailed
shape of ${\bf B}({\bf R})$---for example $B$ can
be essentially constant over a long central region
and then become rapidly larger over a short end region.
(This is the case for the octupole focusing in the
all-electric lattice which is the subject of the
body of the present paper.) 

In any case an adiabatic invariant $I_{\parallel}$
for this motion can be calculated (on-axis) by
\begin{equation}
I_{\parallel} 
= 
\frac{1}{2\pi}\,\oint {\bf P}_{\parallel}\cdot\hat{\bf y}\,dy
=
\frac{m}{2\pi}\,\oint u_{\parallel}dy ,
\label{eq:MagTrap.14}
\end{equation}
Then the period of oscillation can be calculated from
\begin{equation}
T_{\parallel}
=
2\pi\,\frac{\partial I_{\parallel}}{\partial h}.
\label{eq:MagTrap.15}
\end{equation}

\end{document}